\documentclass[11pt,a4paper]{article}
\usepackage{jheppub}
\usepackage{siunitx}
\usepackage{pdfpages}
\usepackage{float}
\usepackage{textcomp}               
\usepackage{braket,slashed,bm}
\usepackage{array,multirow}
\usepackage[normalem]{ulem}
\usepackage{cancel,youngtab}
\usepackage[usenames,dvipsnames]{xcolor}

\usepackage[T1]{fontenc} % if needed
\usepackage{mathrsfs}
\usepackage{booktabs}
\usepackage{adjustbox}
\usepackage{mathtools}
\usepackage{hhline}
\usepackage{soul}

\usepackage{stmaryrd}
\usepackage{trimclip}

\makeatletter
\DeclareRobustCommand{\shortto}{%
  \mathrel{\mathpalette\short@to\relax}%
}

\newcommand{\short@to}[2]{%
  \mkern2mu
  \clipbox{{.5\width} 0 0 0}{$\m@th#1\vphantom{+}{\shortrightarrow}$}%
  }
\makeatother

\usepackage{pgfplots}
\pgfplotsset{compat=1.18}

\usepackage[compat = 1.1.0]{tikz-feynman}
\usepackage{tikz}

\usepackage[compat=1.1.0]{tikz-feynman}

\usepackage{tikz}
\usetikzlibrary{arrows,decorations.pathmorphing,backgrounds,positioning,fit,petri,automata,shadows,calendar,mindmap,
decorations.markings,calc}

\definecolor{labelkey}{rgb}{0,0.5,0.0}

\usepackage{xspace}
\usepackage{slashed}
\usepackage{array,multirow}
\usepackage{graphicx}
\usepackage{graphics}
\usepackage{epsfig}
\usepackage{amsfonts}
\usepackage{amsmath}
\usepackage{amssymb}
\usepackage{dsfont}
\usepackage{color}
\usepackage{pdflscape}
\usepackage{pifont}
\usepackage{pgfplots}
\usepackage{tikz} 
\usepackage{pgfplotstable}
\usepackage{bbold}
\usepackage[capitalise, english]{cleveref}
\usepackage[format=plain,justification=centerlast]{subcaption}
\usepackage{adjustbox}

\newcommand{\lam}{\lambda}

\usepackage{bm}                                  

\newcommand{\beq}{\begin{equation}}
\newcommand{\eeq}{\end{equation}}
\newcommand{\be}{\begin{equation}}
\newcommand{\ee}{\end{equation}}
\newcommand{\bea}{\begin{eqnarray}}
\newcommand{\eea}{\end{eqnarray}}
\newcommand{\ben}{\begin{eqnarray*}}
\newcommand{\een}{\end{eqnarray*}}

\newcommand{\bma}{\begin{pmatrix}}
\newcommand{\ema}{\end{pmatrix}}

\def\lixo#1{}

\def\slashchar#1{\setbox0=\hbox{$#1$}           % set a box for#1
  \dimen0=\wd0                                    % and get its size
  \setbox1=\hbox{/} \dimen1=\wd1                  % get size of/
  \ifdim\dimen0>\dimen1                           % #1 is bigger
    \rlap{\hbox to \dimen0{\hfil/\hfil}}            % so center / in box
    #1                                             % and print #1
  \else                                          % / is bigger
    \rlap{\hbox to \dimen1{\hfil$#1$\hfil}}        % so center #1
    /                                           % and print/
 \fi}                                           %

\newcommand\lsim{\lesssim}

\newcommand{\dslash}[1]{#1 \llap{/\kern-0.5pt}}
\newcommand{\Dslash}[1]{#1 \llap{/\kern+1.5pt}}
\newcommand{\DDslash}[1]{#1 \llap{/\kern+2.3pt}}
\newcommand{\dslashh}[1]{#1 \llap{/\kern+1pt}}

\definecolor{ForestGreen}{rgb}{0.0, 0.27, 0.13}

	\preprint{\begin{flushright} BONN-TH-2024-08
	\end{flushright}}	
	\title{A Novel Proton Decay Signature at DUNE, JUNO, and Hyper-K		}
	
       \author[a]{Florian Domingo,}
        \emailAdd{domingo@physik.uni-bonn.de}
        \affiliation[a]{Bethe Center for Theoretical Physics \& Physikalisches Institut der Universit\"at Bonn,\\ Nu{\ss}allee 12, 53115 Bonn, Germany}
        
		\author[a]{Herbi~K.~Dreiner,}
		\emailAdd{dreiner@uni-bonn.de}
		\author[a]{Dominik K\"ohler,}
		\emailAdd{koehler@physik.uni-bonn.de}

		\author[a]{Saurabh Nangia,}
		\emailAdd{nangia@physik.uni-bonn.de}

		\author[a]{and Apoorva Shah}
		\emailAdd{ashah@uni-bonn.de}

\abstract{

Proton decay, although unobserved so far, is a natural expectation when attempting to explain the baryon asymmetry of the universe.
$p\to K^+\bar{\nu}$ or $p\to K^+\tilde{\chi}_1^0$, with $\tilde{\chi}_1^0$ a light exotic neutral particle, represent possible decay 
channels achievable in models of physics beyond the Standard Model, such as the MSSM with trilinear R-parity-violating terms, or the Standard Model extended by a heavy neutral lepton. Among 
the decay products of these modes, the neutral fermions would typically appear as missing energy in collider searches. The present 
study considers how such decay modes could be differentiated in experimental settings, as the exotic $\tilde{\chi}_1^0$ may further 
decay if it is not protected by a symmetry (such as R-parity in the MSSM). We assess the detection prospects of the proposed 
experiments \texttt{DUNE}, \texttt{JUNO} and \texttt{Hyper-K} in this context. 
}

\begin{document}
\maketitle

%!TEX root= ../RPV_Neutralino_Decay.tex

\section{Introduction}\label{sec:introduction}

Despite its many successes, the Standard Model (SM) of particle physics 
cannot be viewed as an exhaustive description of Nature. One of the 
essential puzzles from the cosmological perspective is embodied by the 
prevalence of matter over antimatter in the observable 
Universe~\cite{Steigman:1976ev, Canetti:2012zc}. 
Baryogenesis~\cite{Sakharov:1967dj} appears as a viable explanation of 
this obvious discrepancy between SM and observations, provided that 
baryon number (B), in particular, is violated in particle interactions. In
the SM, this quantity is accidentally conserved at the perturbative 
level, along with lepton number (L); here, `accidental' means that the most 
general renormalizable gauge-invariant (perturbative) interactions 
automatically preserve these quantum numbers. Violation of the baryon 
number through non-perturbative effects in the SM, such as instantons 
and sphalerons~\cite{tHooft:1976snw, Manton:1983nd, Harvey:1990qw,  
Dreiner:1992vm}, proves insufficient to account for the oberved 
asymmetry\cite{Kuzmin:1985mm, Shaposhnikov:1987tw, Gavela:1993ts, 
Gavela:1994dt}.

Still, theoretical frameworks extending beyond the SM, such as Grand 
Unified Theories (GUTs)~\cite{Georgi:1974sy}, supersymmetry 
(SUSY)~\cite{Wess:1974tw, Wess:1973kz, Haag:1974qh, Nilles:1983ge, 
Haber:1984rc, Martin:1997ns, Dreiner:2023yus}, or theories of quantum
gravity~\cite{Schwarz:1982jn, Green:1982ct}, can effortlessly provide new 
sources of baryon-number violation (or of lepton-number violation when 
considering the leptogenesis mechanism~\cite{Fukugita:1986hr,
Davidson:2002qv,Buchmuller:2004nz,Davidson:2008bu}). However, a drastic 
consequence of introducing such effects is the simultaneous 
emergence of proton decay~\cite{Weinberg:1981wj,Sakai:1981pk,Hawking:1979hw, Antoniadis:1989zy}, the latter remaining unobserved to this date, with the current strongest bound on the lifetime $\tau({p\rightarrow \pi^0 + e^+}) >\SI{1.6e34}{yrs}$~\cite{Super-Kamiokande:2020wjk}. Nevertheless, the host of upcoming next-generation detectors under construction or planned, such as \texttt{DUNE}~\cite{DUNE:2020lwj, DUNE:2020mra, DUNE:2020txw, DUNE:2020ypp}, \texttt{JUNO}~\cite{JUNO:2015sjr, JUNO:2015zny}, 
and \texttt{Hyper-K}~\cite{Hyper-Kamiokande:2018ofw}, 
will certainly help shed a new light on the hypothetical proton instability, and we believe it essential, on the phenomenological side,
to point at possible decay signatures that would be overlooked in the traditional approach.

In this paper, we focus on a proton desintegration dominated by kaons in 
the final states, with modes such as $p\to K^+\bar{\nu}$ and $p\to K^+
\tilde{\chi}_1^0$, where $\tilde{\chi}^0_1$ is a light exotic neutral 
particle with
\begin{equation}
 	m_{\tilde{\chi}^0_1} \leq m_{p}-m_{K^+}\approx 
 	\SI{445}{\mega\electronvolt}\,.
 	\label{eq:chimass}
 \end{equation}
This scenario can be easily embedded within the phenomenology of the 
Minimal Supersymmetric Standard Model with R-parity-violating terms (RPV-MSSM), 
and we shall employ this model as a predictive  framework for our study. However, 
one could also consider instead the decay to a heavy neutral lepton ($N_R$): 
$p\to K^+N_R$, see our \cref{app:HNL}, and for example Ref.~\cite{Fridell:2023tpb}.

Originally motivated by the Hierarchy Problem~\cite{Gildener:1976ai,Veltman:1980mj}, SUSY 
extensions of the SM~\cite{Wess:1974tw} have far-reaching 
phenomenological consequences at low-energy, with features such as dark 
matter candidates, gauge-coupling unification, or neutrino-mass 
generation. We refer the reader to e.g. Refs.~\cite{Nilles:1983ge, 
Martin:1997ns, Barbier:2004ez} for an overview. In contrast to the SM, 
baryon- and lepton-number  violating interactions are naturally 
present in these models at the renormalizable level, unless an additional 
discrete symmetry, often (but not imperatively) 
R-parity~\cite{Farrar:1978xj}, is explicitly 
requested~\cite{Weinberg:1981wj, Martin:1997ns, Dreiner:1997uz, 
Allanach:2003eb}. There exists no deep theoretical motivation to exclude 
such terms via R-parity~\cite{Ibanez:1991pr,Dreiner:2005rd, 
Dreiner:2012ae}, but experimental evidence for a generally B- and L-conserving phenomenology at low-energy implies that they remain 
comparatively suppressed. In particular, they may trigger proton decay, 
which is the feature in which we are interested here. Nevertheless, we 
stress that R-parity conservation does not forbid B- and L-violation in 
higher-order operators, so that proton instability can also be expected 
in the R-parity conserving MSSM when the latter is considered as an 
effective field theory~\cite{Weinberg:1981wj, Ibanez:1991hv, 
Ibanez:1991pr}.

Numerous aspects of proton disintegration in the RPV-MSSM have been 
studied in the literature in the past: the reader may consult 
e.g.~Refs.~\cite{Smirnov:1996bg, Barbier:2004ez, Dudas:2019gkj, 
Chamoun:2020aft}. Two-body final states with only SM particles imply a 
violation of both B and L: spin-conservation indeed dictates the presence 
of an odd number of fermions among the decay products and, in the SM, 
these can only be leptons, then produced in association with a meson. 
Sfermions mediate such processes at tree-level in the 
RPV-MSSM~\cite{Hinchliffe:1992ad,Vissani:1995hp}, provided both B- and L-
violating couplings are simultaneously present. More involved topologies 
have also been considered~\cite{Hall:1983id, Carlson:1995ji,
Long:1997lje,Bhattacharyya:1998bx,Bhattacharyya:1998dt}, leading to bounds 
on a wide set of products of RPV couplings. Nevertheless, proton 
disintegration involving only B-violation is a viable scenario as well, as 
long as a light exotic fermion $\tilde{\chi}_1^0$ is available in the 
spectrum. In particular, the cases involving a light 
photino~\cite{Hall:1983id, Chang:1996sw}, a light gravitino or
axino~\cite{Choi:1996nk, Choi:1998ak} have been considered in the 
literature. In this paper, we identify $\tilde{\chi}_1^0$ with the 
lightest, bino-like neutralino, which is the only phenomenologically 
viable candidate within the strict limits of the MSSM spectrum. A low-mass 
particle of this nature indeed evades existing experimental 
constraints~\cite{Choudhury:1999tn,Dreiner:2009ic}. Astrophysical constraints are also 
satisfied~\cite{Dreiner:2003wh, Dreiner:2013tja}.
Cosmological and astrophysical limits however demand that it is unstable, 
which, in view of the mass and suppressed couplings of this particle, 
typically makes it long-lived. We assume that the bino decays are 
dominated by L-violating channels, triggering final states with leptons and 
mesons. More exotic decay channels involving e.g.~axinos or gravitinos are 
an alternative. Searches of the light neutralino at colliders have been 
recently proposed in~\cite{deVries:2015mfw,Dercks:2018eua, Dercks:2018wum, 
Dreiner:2020qbi,Dreiner:2022swd, Dreiner:2023gir, Gunther:2023vmz}, in the 
more general context of searches for long-lived 
particles~\cite{Curtin:2018mvb, Feng:2022inv,  Berryman:2019dme,  
Lee:2018pag, Filimonova:2019tuy,Alimena:2019zri, DeVries:2020jbs}.

$p\to K^+\tilde{\chi}_1^0$ appears as the simplest proton 
decay mode violating only B in the context of the RPV-MSSM. On a 
superficial level, it shares many similarities with the canonical mode 
$p\to K^+\bar{\nu}$, for which \texttt{Super-Kamiokande} provides the 
(currently strictest) limit $\tau_{p\rightarrow K^++\bar{\nu}}> \SI{5.9
e33}{yrs}$~\cite{Super-Kamiokande:2014otb}. The light long-lived bino 
might distinguish itself from the neutrino through its mass, reducing the 
kaon momentum in experiments. In addition, for decay lengths comparable 
to the detector size, the neutralino could decay within the detector, 
leading to a distinct signature. We analyze both possibilities in detail 
and reinterpret the \texttt{Super-K} bound in this context. We also 
calculate the sensitivities of the upcoming experiments \texttt{DUNE}, 
\texttt{JUNO}, and \texttt{Hyper-K} to such a proton-decay mode.

The paper is organized as follows: in~\cref{sec:model}, we introduce the RPV-MSSM, as well as 
the light-neutralino scenario. We discuss proton decay modes in such a setup, including the 
resulting signatures.
In~\cref{sec:experiment}, we briefly discuss the experimental setups at present and upcoming 
proton-decay search facilities: \texttt{Super-K}~\cite{Super-Kamiokande:2002weg}, \texttt{Hyper-K}, 
\texttt{DUNE} and \texttt{JUNO}. We further discuss the prospects for detecting a light exotic fermion, such as
a neutralino in these detectors, as well as what happens when a proton decays inside a nucleus with large mass number $A$. We conclude this section with a description of
our simulation procedure for 
estimating the sensitivities at the above experiments.
In \cref{sec:Numerical-Analysis}, we outline our numerical 
analysis and present our results. We conclude in~\cref{sec:conclusions}. In~\cref{app:HNL},
we discuss the close connection between the light neutralino we have discussed extensively here and the related heavy neutral lepton scenario.

%!TEX root= ../RPV_Neutralino_Decay.tex
\section{The Signature: Proton Decay followed by Neutralino Decay}
\label{sec:model}
We first discuss the decay of the proton in our model and subsequently the various decays of the light neutralino.

\subsection{Proton Decay to a light Neutralino in the RPV-MSSM}
\label{sec:proton-decay}
As explained in the introduction, we work, for 
convenience, within the framework of the RPV-MSSM model with a light 
neutralino. The renormalizable superpotential may be expressed in 
the notation of Ref.~\cite{Allanach:2003eb}:
\begin{equation}
W = W_{\mathrm{MSSM}} + W_{\mathrm{LNV}} + W_{\mathrm{BNV}}\,,
\label{eq:TFeq1bis}
\end{equation}
where $W_{\mathrm{MSSM}}$ is the MSSM superpotential, and
%$W_{\mathrm{LNV}}$
%contains the lepton number violating interactions and the operators in 
\begin{eqnarray}
W_{\mathrm{LNV}} &=&\epsilon_{ab}\left(\frac{1}{2}\lam_{ijk}L^{ai}L^{bj}
\Bar{E}^k + \lam'_{ijk}L^{ai}Q^{bj}\Bar{D}^k \right)\,, 
\label{eq:RPV-L}\\
W_{\mathrm{BNV}} &= &
\frac{1}{2}\varepsilon_{\alpha\beta\gamma} \lam''_{ijk}\Bar{U}^{\alpha i}\Bar{D}^{\beta j}\Bar{D}^{\gamma k}\,,
\label{eq:TFeq1abis}
\end{eqnarray}
violate lepton- and baryon-number, respectively. In $W_{\mathrm{LNV}}$
we have dropped the bilinear term, as it can be rotated away at a fixed
energy scale \cite{Hall:1983id, Dreiner:2003hw}. $\epsilon_{ab}$ is the
two-dimensional Levi-Civita symbol, and the Latin indices $a,b\in\{1,2\}$ 
are the $\mathrm{SU}(2)_{\text{L}}$ gauge indices in the fundamental 
representation. $\varepsilon_{\alpha\beta\gamma}$ is the three-dimensional 
Levi-Civita symbol and the Greek indices $\alpha,\beta$, and $\gamma\in 
\{1,2,3\}$ denote the $\mathrm{SU}(3)_{\text{C}}$ gauge color indices. 
$\lam''_{ijk}$ is a dimensionless Yukawa coupling and $i, j, k\in\{1, 2, 
3\}$ denote the generation indices. We employ the summation convention. 
$L^i$ and $Q^i$ are the $\mathrm{SU}(2)$-doublet lepton and quark chiral 
superfields; $\bar E^i$, $\Bar{U}^i,$ and $\Bar{D}^i$ are 
$\mathrm{SU}(2)$-singlet electron-, up- and down-type quark chiral superfields, respectively. 

The bino, which we will consider further below, is the spin-1/2 SUSY partner of the 
hypercharge gauge field. As such, its coupling to matter are dictated by the gauge 
interaction. As reminded in the introduction, such a particle might exist with an 
almost arbitrarily low mass without violating collider, astrophysical and 
cosmological constraints, as long as it is unstable.

In the RPV-MSSM, as well as in any viable model of new physics, B- (and 
L-) violation manifests itself at low-energy (the proton mass-scale) 
through the mediation of high-energy fields (taking mass at the 
SUSY-breaking scale, comparable to or larger than the electroweak scale). 
In such configurations with a sizable hierarchy of scales, it is always 
desirable to disentangle long- and short-distance effects through the 
definition of a low-energy effective field theory (EFT), allowing for the resumation of large logarithmic corrections. The impact of new-physics for low-energy particles is 
then summarized within contributions to operators of higher-dimension built out of the light fields. The operators of lowest dimension relevant 
for nucleon decay are of dimension 6: we refer the reader to Ref.~\cite{Chamoun:2020aft} for a recent overview of this specific EFT and 
associated calculation techniques.

The operators that are relevant for the production of a neutral SU(2)$_L$-
singlet fermion $\tilde{\chi}_1^0$ (\textit{i.e.}~neglecting contributions that 
require an electroweak vacuum expectation value, which receive further 
mass-suppression from new-physics) read:
\begin{align}
\hat{\mathcal{Q}}_1^{\prime}
&=\varepsilon_{\alpha\beta\gamma}[(\overline{d^c})^{\alpha}P_R u^{\beta}][(\overline{s^c})^{\gamma}P_R {\tilde{\chi}^{0c}_1}]\,, \notag \\[1.5mm]
\hat{\mathcal{Q}}_1 &=\varepsilon_{\alpha\beta\gamma}[(\overline{s^c})^{\alpha}P_R u^{\beta}][(\overline{d^c})^{\gamma}P_R {\tilde{\chi}^{0c}_1}]\,, \label{eq:chi_operators2}\\[1.5mm]
\hat{\mathcal{Q}}_2 &=\varepsilon_{\alpha\beta\gamma}[(\overline{s^c})^{\alpha}P_R d^{\beta}][(\overline{u^c})^{\gamma}P_R {\tilde{\chi}^{0c}_1}]\,, \notag
\end{align}
where $P_R$ is the right-handed projection operator. $u,d,s$ are the 
four-component quark spinors with $^c$ denoting charge conjugation. 
$\tilde\chi^0_1$ denotes the 4-component neutralino spinor, or any other
exotic neutral fermion.

\begin{figure}
\centering
\begin{subfigure}{.5\textwidth}
  \centering
\includegraphics{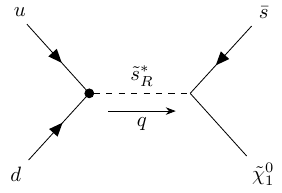}
  \caption{s-channel}
  \label{fig:sub1}
\end{subfigure}%
\begin{subfigure}{.5\textwidth}
  \centering
  \includegraphics{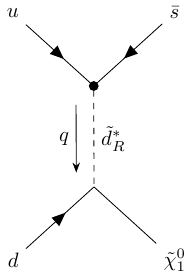}
  \caption{t-channel}
  \label{fig:sub2}
\end{subfigure}
\begin{subfigure}{.5\textwidth}
  \centering
  \includegraphics{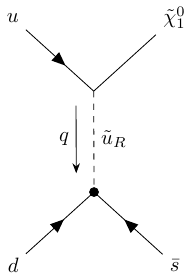}
  \caption{u-channel}
  \label{fig:sub3}
\end{subfigure}
\caption{Feynman diagrams depicting the contributions to the four-fermion operators, which in turn contribute to proton decay. The solid black circle shows the RPV vertex.}
\label{fig:neutralino}
\end{figure}

Contributions to the operators of Eq.(\ref{eq:chi_operators2}) emerge at tree-level in the RPV-MSSM through the mediation of a squark, provided a 
B-violating trilinear coupling $\lambda^{\prime\prime}_{112}=-\lambda^{\prime\prime}_{121}$ is non-vanishing in the superpotential. The 
corresponding Wilson coefficients can be obtained by matching of the EFT and read:
\begin{equation}
\begin{split}
C_{\hat{\mathcal{Q}}'_1}(\mu_R) & = \eta_{\text{QCD}}(\mu_R)\frac{(-\lambda''^*_{112})(\sqrt{2}g')}{3m^2_{\tilde{s}_R}}\,,\\[1.5mm]
C_{\hat{\mathcal{Q}}_1}(\mu_R) & = \eta_{\text{QCD}}(\mu_R)\frac{(-\lambda''^*_{121})(\sqrt{2}g')}{3m^2_{\tilde{d}_R}}\,,\\[1.5mm]
C_{\hat{\mathcal{Q}}_2}(\mu_R) & = \eta_{\text{QCD}}(\mu_R)\frac{(\lambda''^*_{121})(2\sqrt{2}g')}{3m^2_{\tilde{u}_R}}\,,\\
\end{split}
\label{eq:wilsoncoefficients2}
\end{equation}
where $g'$ is the $U(1)_{\text{Y}}$ gauge coupling and the masses in the 
denominator refer to scalar SUSY partners of the right-handed quarks 
mediating the transition: see Fig.\ref{fig:neutralino}. $\eta_{\text{QCD}}
(\mu_R)$ accounts for the renormalization group evolution of the operators 
driven by the strong interaction: its expression can be found in Eq.(6) of 
Ref.~\cite{Chamoun:2020aft}. $\eta_{\text{QCD}}(2\,\mbox{GeV})\approx1.4$.

Naturally, the difficult step consists in extracting hadronic matrix 
elements from these partonic operators. For any operator $\Omega
\in\{\hat{\mathcal{Q}}_1, \hat{\mathcal{Q}}'_1,\hat{\mathcal{Q}}_2
\}$ of Eq.(\ref{eq:chi_operators2}), one can derive the following general 
form~\cite{JLQCD:1999dld} (we employ the $+---$ metric),
\begin{align}
     \langle K^+, \tilde{\chi}^0_1(q) |\Omega|p \rangle &= \bar{v}_{\tilde{\chi}^0_1} P_R \left[W_{0 ,\Omega}^{p \rightarrow K^+}(q^2) - \frac{\slashed{q}}{m_p}W_{1 ,\Omega}^{p \rightarrow K^+}(q^2)\right]u_p\,,\notag\\
     &\equiv \bar{v}_{\tilde{\chi}^0_1} P_R \left[W_{\chi ,\Omega}^{p \rightarrow K^+}(q^2)\right]u_p\,,
     \label{eq:formfactors2}
\end{align}
where $W_{0 ,\Omega}^{p \rightarrow K^+}(q^2)$ and $W_{1 ,\Omega}^{p\to
K^+}(q^2)$ represent the form factors associated with the operator 
$\Omega$ for the transition $p \rightarrow K^+$, depending on the 
squared momentum-transfer, $q^2$ ($=m^2_{\tilde{\chi}_1^0}$ in our 
case). $u_p$ and $v_{\tilde{\chi}_1^0}$ denote the four-component 
spinors for the proton and neutralino respectively.

Lattice evaluations of the $p\to K^+$ 
form-factors~\cite{JLQCD:1999dld,Aoki:2017puj, Yoo:2021gql} focus on 
the limit $q^2\to0$ (hence on $W_{0,\Omega}^{p\to K^+}$), corresponding 
to the neutrino final-state. This limit does not necessarily apply in 
the case of a neutralino. Alternatively, it is possible to compute the 
form factors in chiral perturbation theory and determine chiral 
parameters from the lattice \cite{JLQCD:1999dld}, thus retaining full 
momentum dependence (at leading order in chiral perturbation theory, in 
practice).

The decay amplitude for the proton decay may now be expressed as follows:
\begin{equation}
\begin{split}
\mathcal{A}\left(p \rightarrow K^+ + \tilde{\chi}^0_1 \right) & = i\sum_{\Omega}C_{\Omega}\bar{v}_{\tilde{\chi}^0_1} \left\{ P_R \left[W_{0 ,\Omega}^{p \rightarrow K^+}(q^2)\right] + P_L\left[\tfrac{m_{\tilde{\chi}^0_1}}{m_p}W_{1 ,\Omega}^{p \rightarrow K^+}(q^2)\right]\right\}u_p\,,
\end{split}
\end{equation}
leading to the partial decay width:
\begin{align}
    &\Gamma \left(p \rightarrow K^+ + \tilde{\chi}^0_1 \right) = \frac{m_p}{32\pi}\Bigg[1-2\frac{m_K^2+m_{\tilde{\chi}^0_1}^2}{m_p^2}+\left(\frac{m_K^2-m_{\tilde{\chi}^0_1}^2}{m_p^2}\right)^2\Bigg]^{1/2}\label{eq:protonwidth}
    \\[1.5mm]
    &\null\hspace{1.4cm}\Bigg\{\left(1+\frac{m_K^2-m_{\tilde{\chi}^0_1}^2}{m_p^2}\right)\left[ \left|\sum_{\Omega}C_{\Omega} W_{0 ,\Omega}^{p \rightarrow K^+}(m_{\tilde{\chi}^0_1}^2)\right|^2+\frac{m_{\tilde{\chi}^0_1}^2}{m_p^2}
    \left|\sum_{\Omega}C_{\Omega} W_{1 ,\Omega}^{p \rightarrow K^+}(m_{\tilde{\chi}^0_1}^2)\right|^2\right]\nonumber\\[1.5mm]
    &\null\hspace{3.8cm}+4\frac{m_{\tilde{\chi}^0_1}^2}{m_p^2}\mbox{Re}\left[\left(\sum_{\Omega}C_{\Omega} W_{0 ,\Omega}^{p \rightarrow K^+}(m_{\tilde{\chi}^0_1}^2)\right)^*\sum_{\Omega}C_{\Omega} W_{1 ,\Omega}^{p \rightarrow K^+}(m_{\tilde{\chi}^0_1}^2)\right]\Bigg\}\,.\nonumber
\end{align}

\begin{figure}
    \centering
    \includegraphics[scale=0.5]{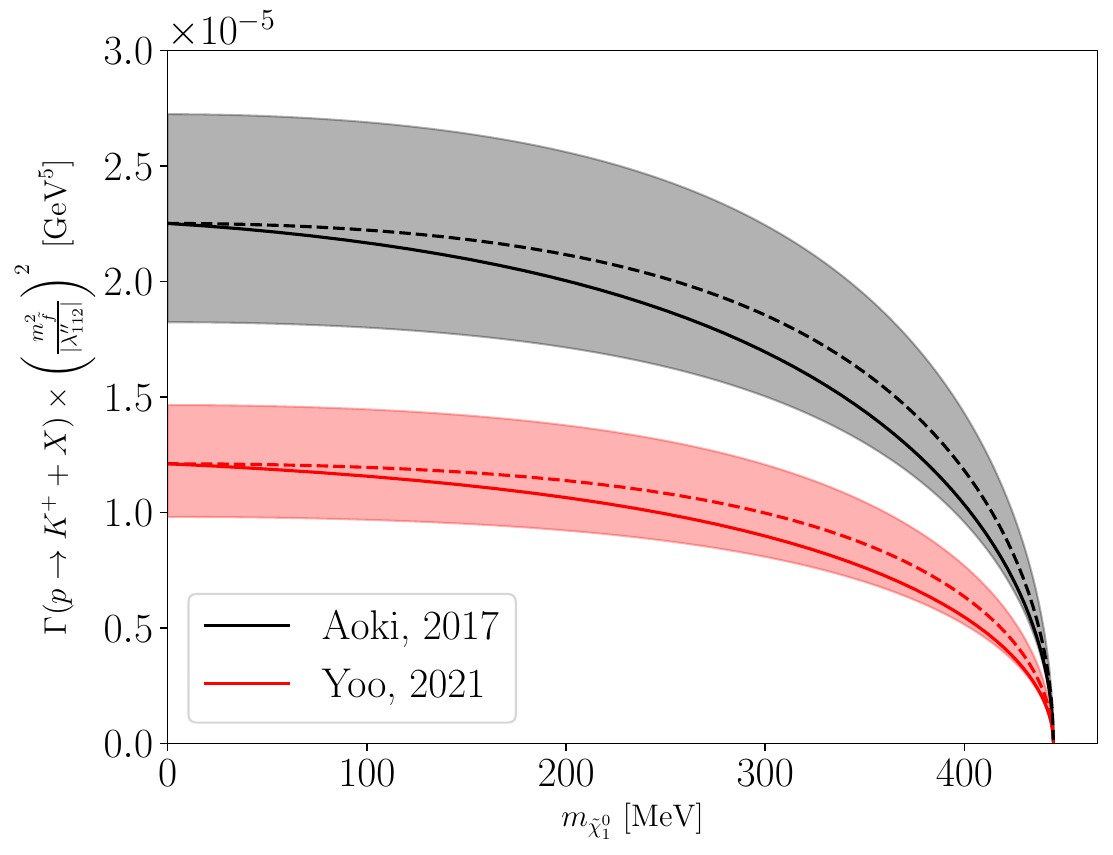}
    \caption{Proton decay width normalised to $|\lam^{\prime\prime}_{112}|^2/m_{\tilde{f}}^4$,
where $m_{\tilde{f}}$ represents a universal value for the squark masses. Two different lattice evaluations are used for the numerical values of the form factors: from Aoki, 2017~\cite{Aoki:2017puj} and 
Yoo, 2021~\cite{Yoo:2021gql}. The dashed line represents the case where lattice form-factors 
at $q^2=0$ are used, whatever the neutralino mass, while the solid lines represent results with form factors determined according to chiral perturbation theory~\cite{JLQCD:1999dld}. The bands around the dashed lines denote the approximate error in the lattice calculation of the form factors.}
    \label{fig:proton_decay_width_change}
\end{figure}

The decay width normalized to $|\lam^{\prime\prime}_{112}|^2/m_{\tilde{f}}^4$,
where $m_{\tilde{f}}$ represents a universal value for the squark masses, is 
shown in \cref{fig:proton_decay_width_change}
for various lattice inputs. Here, we use the results of the two lattice approaches presented in Refs~\cite{Aoki:2017puj,Yoo:2021gql}. The solid lines employ the form-factors retaining full momentum-dependence, while the dashed lines have been obtained under the approximation $W_{\chi,\Omega}^{p\to K^+}(q^2)\approx W_{\chi,\Omega}^{p\to K^+}(0)$. The dotted lines account for the quoted lattice uncertainties at $1\sigma$. We observe that the momentum dependence in the form-factors can 
be neglected in view of the large uncertainty originating in the 
lattice modelization. Obviously, no better than an order of magnitude 
can be set on the actual size of the hadronic matrix elements, while the 
variations due to momentum dependence typically remain under 10\%.

Assuming that the neutralinos would behave as an invisible particle at \texttt{Super-K}, one may exploit the limit of this experiment for the 
decay rate $p\to K^+\bar{\nu}$. In the massless neutralino limit, this led Ref.~\cite{Chamoun:2020aft} to the limit:
\begin{equation}
    |\lambda''_{121}|/m^2_{\tilde{f}} <  3.9\times 10^{-31}\,\SI{}{\giga\electronvolt}^{-2}\,,
\label{eq:flodobound2}
\end{equation} 
updating older estimates. This bound can naively be extended to massive 
neutralinos through a rescaling by the square-root of the the quantity depicted in 
\cref{fig:proton_decay_width_change}.

Nevertheless, this simplistic picture holds only under the 
approximation where the modified kaon kinematics (due to the neutralino 
mass) and subsequent neutralino decays have a negligible impact on the 
experimental strategy. The purpose of this paper exactly consists in 
demonstrating how these two effects may leave their imprint on the 
experimental results, allowing, under favorable conditions, to 
disentangle the decay mode involving the neutralino from the more 
classical channels with a(n anti)neutrino in the final state.

\begin{figure}
    \centering
    \includegraphics[scale=0.5]{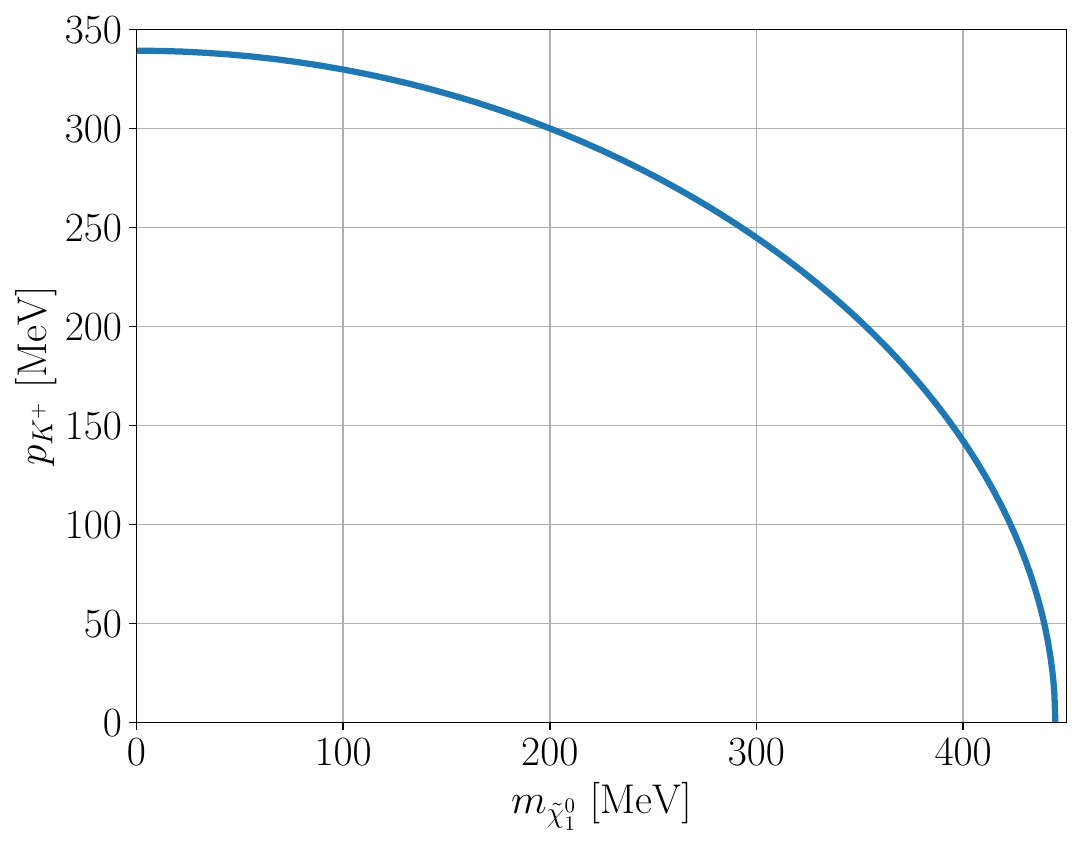}
    \caption{Momentum of the final state kaon as a function of the neutralino mass in the process $p \rightarrow K^+ + \tilde{\chi}^0_1$.}.
    \label{fig:kaonmomentum}
\end{figure}

Considering the kinematical effect first, we write the kaon momentum in the rest-frame of the proton:
\begin{align}
|\vec{p}_{{K^+}}|= \frac{m_p}{2} \sqrt{1-2\frac{m_K^2+m_{\tilde{\chi}^0_1}^2}{m_p^2}+\left(\frac{m_K^2-m_{\tilde{\chi}^0_1}^2}{m_p^2}\right)^2}\,.
\label{eq:kaonmomentum2}
\end{align}
This quantity is shown in Fig.~\ref{fig:kaonmomentum}. In the massless neutralino limit (or in the case of a neutrino in the final state), $|\vec{p}_{{K^+}}| \approx \SI{339}{\MeV}$.

\subsection{Neutralino Decay}

The second experimental handle on the decay channel involving the 
neutralino rests with the possible observation of neutralino decay 
products within the detector, which obviously only applies if the 
neutralino is sufficiently short-lived. The decay modes of a light 
neutralino have been recently reviewed in Ref.~\cite{Domingo:2022emr} 
and we briefly present the channels relevant for our study. Under the 
assumption that neutralino decays involving lighter exotic fermions, 
such as axinos or gravitinos, in the final state are absent or 
subdominant, neutralino disintegrations necessarily violate lepton 
number, as the only accessible lighter fermions (needed for the
conservation of angular momentum) are muons, electrons or neutrinos. Interactions of this type are possible in the RPV-MSSM,
see Eq.~(\ref{eq:RPV-L}).

The first possibility is that neutralino decays are controlled by 
$LQ\bar{D}$ operators. Couplings of this type lead to semi-leptonic 
disintegrations. Considering the mass-regime relevant here, $m_{\tilde 
{\chi}^0_1} \lesssim \SI{445}{\mega\electronvolt}$, a pion would be the 
only kinematically accessible meson. The expressions for the partial 
decay widths $\Gamma\left(\tilde{\chi}^0_1\rightarrow\pi^{\pm/0}+\ell 
_i^\mp/\nu_i \right)$ in the limit of a pure bino state can be found in
Ref.~\cite{deVries:2015mfw}. The RPV couplings involved here 
(discarding CKM mixing, see \cite{Dreiner:1991pe, Agashe:1995qm}) are 
$\lam^{\prime}_{i11}$, $i=1,2,3$.\footnote{Note for the decay 
$\tilde{\chi}^0_1 \rightarrow \pi^\pm+\ell^\mp_i$ only $i=1,2$ is 
kinematically possible.}
In addition, we stress that, with both $\lam^{\prime\prime}_{112}$ and 
$\lam^{\prime}_{i11}$ non-vanishing, protons may directly employ the 
decay mode $p\to K^+\bar{\nu}_i$ (see e.g.
Ref.~\cite{Chamoun:2020aft}). For small $\lambda^{\prime}_{i11}$ (which is the relevant regime with a long-lived bino), the associated decay 
width is correspondingly suppressed, however, with respect to the $p\to K^+\tilde{\chi}_1^0$ mode (provided the latter is kinematically open).
Thus, these two proton decay modes can only compete close to the neutralino production threshold, 
or for $\lambda^{\prime}_{i11}\approx1$.

Purely leptonic decays of the neutralino can be mediated by operators 
of the $LL\Bar{E}$-type. In the pure bino approximation, neglecting all 
mixing effects in the sfermion sector and exploiting $m_{\tilde{\chi}
_1^0} \ll m_{\tilde{f}}$, where $m_{\tilde{f}}$ represents a universal 
sfermion mass, one obtains~\cite{Dreiner:2008tw}:
\begin{align}
   \Gamma\left(\tilde{\chi}^0_1 \rightarrow \bar{\nu}_i + \ell^+_j + \ell^-_k\right) = \frac{3 g^{\prime\hspace{0.04cm} 2} |\lambda_{i j k}|^2}{2^{12} \pi^3} \frac{m_{\tilde{\chi}_1^0}^5 }{m_{\tilde{f}}^4}\,, \quad (i\neq j)\,.
   \label{eq:neutralino-decay-LLE}
\end{align}

Finally, both $LQD$ and $LLE$ operators produce the radiative 
decays~\cite{Hall:1983id,Dawson:1985vr,Haber:1988px, Domingo:2022emr, 
Dreiner:2022swd},
\begin{align}
	\Gamma(\tilde\chi^0_1\to \gamma +\nu_i/\bar{\nu_i})&=
	\frac{\alpha^2m^3_{\tilde{\chi}^0_1}} {512\pi^3\cos^2\theta_W}
 %\sum_f
 |\lam^{(\prime)}_{iff}|^2\left[
	\frac{e_f N_c^f m_f\left(4e_f+1\right)}{m^2_{\tilde{f}}} \left(1+\log{\frac{m^2_f} 
		{m^2_{\tilde{f}}}}\right)\right]^2\,,\label{eq:neutralinoloop2}
\end{align}
where $\lam^{(\prime)}_{iff}$ is the relevant trilinear coupling 
($L_iQ_j\bar D_j$ or $L_iL_j\bar E_j$), $f$/$\tilde{f}$ the associated 
fermions/sfermions running in the one-loop diagram, $e_f$ their 
electric charge in units of $e$, $m_f$/$m_{\tilde{f}}$ their masses, 
$N_c^f$ their color number (1 or 3). Once again, simplifying 
assumptions have been used as to the scalar sector.

Both the tree-level and radiative decay modes can be relevant for 
light neutralinos~\cite{Dreiner:2022swd}. Depending on the neutralino 
mass the tree-level decay modes might be kinematically inaccessible 
while the radiative mode has, essentially, no threshold. Even in 
scenarios where both modes are possible, the radiative decay becomes 
particularly important for very light neutralinos as the mass 
dependence makes evident: $\left(m^3_{\tilde{\chi}^0_1}m^2_f \right)/m^4_{\tilde f}$, 
\textit{cf.}~\cref{eq:neutralinoloop2}, against $m^5_{\tilde{\chi}^0_1}/
m^4_{\tilde f}$ for the tree-level three-body decay into leptons, 
\textit{cf.}~\cref{eq:neutralino-decay-LLE}.

%!TEX root= ../RPV_Neutralino_Decay.tex

\section{Proton Decay Experiments}\label{sec:experiment}
In this section, we present the proposed proton decay experiments that we 
focus on in this study: \texttt{DUNE}, \texttt{JUNO}, and \texttt{Hyper-K}. 
We summarize the important technical features for each detector 
in~\cref{tab:detectorinfo}.

\subsection{Detectors}
\texttt{DUNE}, or the Deep Underground Neutrino 
Experiment~\cite{DUNE:2020lwj, DUNE:2020ypp, DUNE:2020mra, DUNE:2020txw}, 
currently under construction, consists of a far detector (FD), situated 
\SI{1.5}{\kilo\metre} underground at the Sanford Underground Research 
Facility (SURF), about 1300km west of the Fermi National Accelerator 
Laboratory (FNAL). The FD is divided into four cuboidal detector volumes. 
Each has the dimensions \SI{58.2}{\metre}$\times$\SI{14}{\metre}$\times$
\SI{12}{\metre} and consists of modular Liquid Argon Time Projection 
Chambers (LArTPCs) with a fiducial mass of \SI{10}{\kilo\tonne}. Charged 
particles ionize the medium and also produce scintillation light as they 
drift along the chamber. At the planned start of the beam run at FNAL, two
FDs will be deployed, in two separate cryostat chambers. The first detector 
module is scheduled to be operational by 2026. In this work, we 
consider four chambers with a combined fiducial volume of \SI{40}{\kilo 
\tonne} to derive our results, and assume that proton decays in one chamber 
are searched for in the same chamber, \textit{i.e.} we do not consider the 
possibility that decays from one chamber are detected in 
another\footnote{The information on the spacing between the detectors and 
their arrangement is not clearly presented in the TDRs. Thus a combined 
analysis is currently not possible.}. Prototype studies  estimate the 
detection efficiency at $30\%$ for $p\to K^++ \bar{\nu}
$~\cite{DUNE:2020ypp}. The LArTPC technology can, in principle, reduce the 
background below single-event level for key nucleon decay 
channels~\cite{DUNE:2020lwj}.

The Jiangmen Underground Neutrino Observatory, or 
\texttt{JUNO}~\cite{JUNO:2015sjr, JUNO:2015zny}, is a proposed spherical 
liquid scintillator detector of inner diameter \SI{35.4}{\metre} with \SI{20}
{\kilo\tonne} fiducial mass, situated \SI{700}{\metre} underground at 
Kaiping in Southern China. It is expected to start operations in 2024. The 
current choice for the liquid scintillator material is 
linear alkylbenzene. The detector is submerged in a water pool to protect it
from radioactivity from the surrounding rock and air. The expected 
background for $p\to K^+ + \bar{\nu}$ is 0.5 events per 10 years, and the 
expected detection efficiency reaches about $36.9\%$~\cite{JUNO:2022qgr}.

The Hyper-Kamiokande observatory, or 
\texttt{Hyper-K}~\cite{Hyper-Kamiokande:2018ofw}, is the successor to the 
Super-Kamiokande (\texttt{Super-K}) 
experiment~\cite{Super-Kamiokande:2002weg}, and is a proposed large-scale 
underground water Cherenkov neutrino detector in Japan. Its proposed start 
of operations is around the year 2025. It consists of a cylindrical vertical 
tank with a diameter of \SI{74}{\metre} and height of \SI{60}{\metre}. The 
fiducial volume contains highly transparent purified water with a fiducial 
mass of about \SI{187}{\kilo\tonne}. While most of the background is 
scattered by the surrounding water accross the fiducial volume, or rejected 
through veto-detectors, neutron and kaon backgrounds from cosmic rays 
persist. Detection of the process $p\to K^+ + \bar{\nu}$ is achieved  
through the reconstruction of the pionic or muonic decay modes of the kaon. 
The former, owing to lower background, leads to stricter bounds. Its 
efficiency is expected to be about $10.8\pm1.1$\%. The corresponding 
expected background is $0.7 \pm 0.2$ red events per \SI{}{\mega\tonne} per year.

\begin{table}[t]
\renewcommand{\arraystretch}{1.5}

    \centering
   \begin{adjustbox}{max width=\textwidth}
    \begin{tabular}{|c|c|c|c|c|}
    \hline
    
         &\textbf{\texttt{Super-K}}&\textbf{\texttt{Hyper-K}}&\textbf{\texttt{JUNO}}&\textbf{\texttt{DUNE}}  \\
         \hline
         \hline
        Geometry&Cylindrical &Cylindrical&Spherical&Cuboidal (4 modules)\\
        & \SI{42}{\metre} height$\times$\SI{39}{\metre} diameter & \SI{60}{\metre} height$\times$\SI{74}{\metre} diameter& \SI{35.4}{\metre} diameter & \SI{58.2}{\metre}$\times$\SI{14}{\metre}$\times$\SI{12}{\metre} \\
        \hline
        Detector Material & Water & Water & LABs & Liquid Argon\\
        \hline
        Working Principle & Cherenkov& Cherenkov & Scintillation & Scintillation\\
        \hline
        Fiducial Mass & \SI{22.5}{\kilo\tonne}&\SI{187}{\kilo\tonne}& \SI{20}{\kilo\tonne}&\SI{40}{\kilo\tonne}\\
        \hline
        No. of Protons & $\sim 7.5\times 10^{33}$ & $\sim6.3\times10^{34}$& $\sim6.9\times 10^{33}$ & $\sim1.1\times 10^{34}$\\
        \hline
        $\epsilon_{inv.}$  & $\mathcal{O}\left(10\right)\%$ & $\mathcal{O}\left(10\right)\%$& 37\% & 30\%\\
        \hline
    \end{tabular}
    \end{adjustbox}
    \vspace{0.5cm}
    \caption{Summary of technical details for the upcoming detectors: 
    \texttt{Hyper-K}~\cite{Hyper-Kamiokande:2018ofw}, \texttt{JUNO}~\cite{JUNO:2015sjr}, 
    and \texttt{DUNE}~\cite{DUNE:2020lwj}. We include the characteristics of 
    \texttt{Super-K}~\cite{Super-Kamiokande:2002weg} for comparison. $\epsilon_{inv.}$ is the approximate detection efficiency of the charged kaon.} % for the neutrino (or invisible neutralino) mode.}
    \label{tab:detectorinfo}
\end{table}

%%%%%%%%%%%%%%%%%%%%%%%%%%%%%%%%%%%%%%%%%%%%%%%%%%%%%%
\subsection{Prospects for the Detection of the 
Neutralino Final State}
%%%%%%%%%%%%%%%%%%%%%%%%%%%%%%%%%%%%%%%%%%%%%%%%%%%%%%
\label{subsec:neutralino-detection}
\texttt{Hyper-K} offers a significantly higher fiducial volume, compared to
the other detectors, but it is limited by its detection principle: 
\texttt{Hyper-K} (and \texttt{Super-K}) cannot measure the momentum for 
particles below the Cherenkov threshold of water, which for kaons, is
about $\SI{750}{\mega\electronvolt}
$~\cite{Super-Kamiokande:2014otb}.\footnote{It can detect the kaon through 
its decay products since these are produced above the Cherenkov threshold.} 
This is a serious disadvantage for distinguishing between proton decay modes 
through a measurement of the momentum of the outgoing kaon, since the kaons 
from $p\to K^+ + \bar{\nu}$ are produced with a momentum of about \SI{340}, and the value is even lower in the case of a massive 
neutralino in the decay $p\rightarrow K^++\tilde{\chi}^0_1$, \textit{cf.}
\cref{fig:kaonmomentum}. Thus, \texttt{Hyper-K} cannot distinguish the case 
of a kaon produced in association with a detector-stable (or invisibly 
decaying) massive neutralino from that of a decay involving a neutrino. 
Despite their smaller dimensions, \texttt{DUNE} and \texttt{JUNO} are very competitive in 
this respect: being scintillation detectors, they can measure the kinetic 
energy of the kaon down to a threshold of $\SI{50}{\MeV}$.

Turning to the detection strategy based on measuring the neutralino decay,
we assume zero background and a neutralino detection efficiency, 
$\epsilon_{\text{vis.}}$, of $100\%$ in our simulations. The characteristic 
signature includes visible objects from the kaon decay, as well as from the
neutralino decay, with a likely displaced vertex structure. 
Timing-coincidence effects should provide us with various methods of 
implementing high-efficiency cuts leading to a clean signal-to-background 
ratio. However, the subsequent decay products of the neutralino, e.g. the 
muons and pions, might be invisible to the  \texttt{Super-K} and \texttt{ 
Hyper-K} detectors, again due to the threshold for Cherenkov radiation.\footnote{The Cherenkov threshold for muons is
\SI{160}{\MeV} for the total energy, or \SI{115}{\MeV} 
for the kinetic energy. The threshold for charged pions is 
\SI{210} {\MeV} in the total energy and \SI{160}{\MeV} for the kinetic.} 
The 
momenta of the muons and pions are fully determined by the mass and momentum 
of the parent neutralino. Depending on this momentum distribution, only a small fraction should induce Cherenkov radiation. 
Furthermore, due to the energy loss of the muons or pions, these should 
come to a stop inside the detector volume (in about 2\,m) and produce characteristic subsequent decay products. 

The muon, as well as the pion, can both be identified by a Michel electron. 
Thus, the event can be distinguished from competing processes by the 
identification of its final states. The preceding proton decay can be used 
as a trigger to reject background events. On the other hand, \texttt{DUNE} 
and \texttt{JUNO} can directly identify secondary particles in their 
tracking chambers. A precise estimate of their performance, taking into 
account the exact shape of the fiducial volume, would 
however require detailed simulations for each detector.

\subsection{Proton Decay in Nuclei}

In our discussion so far, we have considered free protons. Yet most of 
the protons in the detector materials are in bound states. \texttt{Super-K} 
and \texttt{Hyper-K} use water (H$_2$O) with most of the protons in the
oxygen nucleus, \texttt{DUNE} uses liquid argon, and \texttt{JUNO} employs 
linear alkylbenzene (C$_6$H$_5$C$_n$H$_{2n+1}$) as proton sources.  We now 
briefly discuss the impact of this distinction for our analysis.

Effects such as Fermi momentum, correlations between nucleons, and nuclear 
binding energy can alter the decaying proton's momentum, and thus the 
kaon momentum given in \cref{eq:kaonmomentum2}, in the lab frame. These 
effects can be analyzed through various approaches -- for instance, by 
modeling the nucleus as a local Fermi gas~\cite{Stefan:2008zi}, or through 
the so-called approximated spectral function approach~\cite{Benhar:2005dj, 
Ankowski:2007uy}. In addition, the produced kaon can re-scatter on 
the surrounding nucleons, further affecting its momentum, as it exits the 
nucleus. This can be modeled as a Bertini cascade~\cite{Bertini:1963zzc, 
Bertini:1969rqi,Bertini:1971xb}, implemented in 
\texttt{GEANT4}~\cite{GEANT4:2002zbu,Heikkinen:2003sc}.

Due to the above effects, the kaon momentum is no longer fixed at the 
value of~\cref{eq:kaonmomentum2}, but smeared out instead. A significant 
fraction of kaons lose energy, although some may also gain energy due to 
collisions with high-momentum nucleons. A detailed discussion for 
\texttt{Super-K} considering the $^{16}$O nucleus can be found in 
Ref.~\cite{Super-Kamiokande:2014otb}. A simulation for the case of argon 
can be found in Ref.~\cite{Stefan:2008zi}. A study for liquid scintillators 
can be found in Ref.~\cite{Undagoitia:2005uu}, according to which we expect 
a comparable behavior with linear alkylbenzene.

For our study, the smearing effects are not expected to be 
of critical importance, since we do not 
attempt to simulate momentum measurements of the kaon, but  focus instead on the search where the kaon is 
 reconstructed through the observation of its decay products. This choice has a minor impact on the detection efficiency. For the momentum measurements, 
 we will directly employ the sensitivity estimates claimed by the \texttt{DUNE} and \texttt{JUNO} colaborations. In the case of 
\texttt{Super-K} and \texttt{Hyper-K}, the kaon reconstruction relies on 
the kaon decaying at rest. Due to the 
effects discussed above, some kaons become 
energetic enough that they do not stop in the water before decaying. 
However, \texttt{Super-K} estimates this number to be only $11\%
$~\cite{Super-Kamiokande:2014otb}. 

In the case of \texttt{DUNE}, kaon tracks with a momentum higher 
than \SI{180}{\MeV} can be measured with a detection efficiency 
of 90\%~\cite{DUNE:2020ypp}. If the kaon momentum is below this 
threshold, the track cannot be reconstructed due to the short 
decay length of the kaon ($\beta\gamma\tau_K$). However, the 
reconstruction algorithms used in the experiment can still
identify the kaon through its dominant decay chain, $K\to\mu\to 
e$, with an efficiency higher than 90\%.\footnote{A study 
performed by \texttt{DUNE} determined the detection efficiency 
for a kaon momentum in the range \SIrange{150}{450}
{\MeV}~\cite{DUNE:2020ypp}. The efficiency can be up to $\sim 85
\%$ for a kaon with a kinetic energy of \SI{200}{\MeV}. For 
kinetic energies below \SI{50}{\MeV} the efficiency is lower
than for the reconstruction of the decay products, \textit{i.e.} 
30\%. This efficiency is not used in our study but is claimed to 
be technically achievable. }

For \texttt{JUNO} the search strategy relies on the observed time 
coincidence and well-defined energies of the kaon decay 
products~\cite{JUNO:2015sjr}. Additionally, the liquid scintillator detects 
a prompt signal coming from the kaon. We assume that the efficiency for kaon 
reconstruction  with this approach is constant for all momenta. A later 
study~\cite{JUNO:2022qgr} indicates that the best reconstruction can be 
achieved by a time-correlated triple coincidence, 

composed of the energy deposit of the kaon, a short-delayed deposit of its decay daughters and the energy deposit of the Michel electron from the muon decay.
A detailed simulation of detector effects and proton momentum distribution for \texttt{DUNE} and \texttt{JUNO} would allow for an enhanced efficiency in the considered proton decay mode. However, this goes beyond the scope of this work.

Thus, we shall neglect any nuclear effects. However, we wish to emphasize that, 
in order to 
determine the mass of the neutralino from the kaon decay, a precise 
momentum measurement of the kaon is required. In that case, an accurate 
description of nuclear effects is crucial.

Before concluding this subsection, we briefly mention one more effect. A 
proton decaying inside a nucleus can leave the nucleus behind in an excited 
state. The latter de-excites promptly into the ground state by the emission
of a gamma ray. One can estimate the energy of this photon and search for it 
as a coincidence signal. With this method, it is 
possible to reduce background events from cosmic ray muons and radioactivity of materials 
around the detector wall. Indeed, \texttt{Super-K} has searched for such gamma 
rays~\cite{Super-Kamiokande:2014otb}, and \texttt{Hyper-K} will be able to as 
well. \texttt{JUNO} can exploit the time-coincidence
between the prompt signal from kaons hitting the liquid scintillator and 
a delayed signal from its decay products, to reduce background events. However, 
\texttt{DUNE} indicates that measuring time differences on the scale of the kaon 
lifetime is difficult~\cite{DUNE:2020ypp}.

%!TEX root= ../RPV_Neutralino_Decay.tex

\subsection{Simulation Procedure}\label{sec:simulation}
We now describe the simulation procedure that we use in order 
to estimate the sensitivities of the various experiments for detecting a 
proton decaying into a neutralino, possibly followed by the decay of the 
neutralino.

The total number of produced neutralinos (or kaons) in a given 
experimental setup may be written as, 
\begin{equation}
    N_{\tilde{\chi}^0_1}^{\text{prod}} = N_p \cdot \Gamma(p \rightarrow K^+ + \tilde{\chi}^0_1) \cdot t\,,
\end{equation}
where $\Gamma(p \rightarrow K^+ + \tilde{\chi}^0_1)$ is 
calculated in~\cref{eq:protonwidth}, $N_p$ is the total number 
of protons in the detector volume and $t$ is the runtime of the 
experiment. If the neutralino is not explicitly looked for, the 
total number of observed events can then be estimated by simply 
multiplying the above expression by the efficiency of kaon 
detection, \textit{cf.} discussion in the previous section. 

In the case where the neutralino may decay visibly into a final state 
$X$, however, we also need to determine the number of such decays that 
can be reconstructed within the detector volume. We estimate them 
as, 
\begin{equation}
    N^{\text{obs.}}_{\tilde{\chi}^0_1 \shortto X} = N_{\tilde{\chi}^0_1}
    ^{\text{prod}} \cdot \langle P[\tilde{\chi}^0_1\ \text{in d.r.}]
    \rangle \cdot \epsilon_{\text{vis.}}\,.
\end{equation}
Here the function $\langle P[\tilde{\chi}^0_1\ \text{in d.r.}]\rangle$ 
represents the average probability 
of the neutralino to decay within the 
fiducial volume or the detectable region (d.r.) of the detector. This probability is dependent on the 
neutralino's lifetime, kinematics, point-of-origin within the detector, 
and the geometry of the detector itself. $\epsilon_{\text{vis.}}$, in the
above, is the detection efficiency for the visible state $X$, which we 
shall ultimately set to $100\%$ in this work, 
\textit{cf.}~\cref{subsec:neutralino-detection}.

In order to estimate $\langle P[\tilde{\chi}^0_1\ \text{in d.r.}]\rangle$ 
for each detector, we run a Monte Carlo simulation with $N^{\text{MC}}_
{\tilde{\chi}^0_1}$ neutralinos of a given mass and with fixed RPV couplings.\footnote{In practice, we set $N^{\text{MC}}_
{\tilde{\chi}^0_1}$ to a value of about 10,000 to 50,000 events.}
The neutralinos originate at random 
points within the detector, and travel in random directions. We 
discuss the geometry details for each detector below. Then, we estimate,
\begin{equation}
\langle P[\tilde{\chi}^0_1\ \text{in d.r.}]\rangle = \frac{1}{N_{\tilde{\chi}^0_1}^{\text{MC}}}\sum_{i=1}^{N_{\tilde{\chi}^0_1}^{\text{MC}}}P_i[\tilde{\chi}^0_1\ \text{in d.r.}]\,,
\end{equation}
where,
\begin{equation}
    P_i[\tilde{\chi}^0_1\ \text{in d.r.}]=1- e^{-L_i/\lambda}\,,
    \label{eq:indivprob}
\end{equation}
is the individual probability for the $i\textsuperscript{th}$ simulated 
neutralino to decay inside the detector. $L_i$ is the distance between 
the point where the neutralino originates and the detector boundary along 
its direction-of-travel. The mean decay length $\lam$ (independent of
$i$) is given by,
\begin{align}
\lambda &= \gamma\beta/\Gamma_{\text{tot}}\,,
\end{align}
with $\Gamma_{\text{tot}}$ the total decay width of the 
neutralino and,
\begin{align}
\gamma = E/m_{\tilde{\chi}^0_1}\,, \qquad
\beta = \sqrt{\gamma^2 -1}/\gamma\,.
\end{align}
In the above, $m_{\tilde{\chi}^0_1}$ and $E$ are the neutralino mass and 
energy, respectively. Thus, $L_i$ is the only geometry-dependent factor. 
We now describe how we calculate it for the considered detectors.

\paragraph{\texttt{Hyper-K}:} \texttt{Hyper-K} is a vertical 
cylindrical-shaped detector with a radius of $R= \SI{37}{\m}$ and height 
$H=\SI{60}{\m}$. Let the $i\textsuperscript{th}$ neutralino be generated 
inside the \texttt{Hyper-K} volume at a point $(r, \varphi, z)$ in a 
cylindrical coordinate system with origin at the center of the bottom 
surface ($z=0$) of the detector. Let its three-velocity, $\vec{v}$, be at
azimuthal angle $\varphi_v$ with components $v_z$ and $v_\perp=\sqrt{ 
{\vec v}^{\,2}-v_z^2}$ along the $z$-axis and in the polar plane, 
respectively. Then, we have,
\begin{equation}
L_i = |\vec{v}| \times \text{min}(t_1, t_2)\,,
\end{equation}
where,
\begin{equation}
    t_1 \equiv \begin{cases}
    {\displaystyle\frac{(H-z)}{v_z}}, & \text{if } v_z > 0\,,\\[3.5mm]
    {\displaystyle-\frac{z}{v_z}}, & \text{if } v_z < 0\,,\\
          \end{cases}
\end{equation}
and $t_1 > t_2$ if $v_z=0$. Here, $t_2$ is given by,
\begin{equation}
t_2 = \frac{-r \text{cos}(\varphi-\varphi_{v}) + \sqrt{R^2 - r^2 \text{sin}^2(\varphi-\varphi_{v})}}{|v_{\perp}|}\,.
\end{equation}
\texttt{Super-K} can be modeled analogously.

\paragraph{\texttt{DUNE}:} \texttt{DUNE} has four cuboid-shaped FDs with
dimensions $L = \SI{58.2}{\m}$, $W = \SI{14.0}{\m}$, and $H= \SI{12.0}
{\m}$. In a rectangular coordinate system with the origin at the bottom corner of the detector, for the 
$i\textsuperscript{th}$ neutralino generated at $(x, y, z)$ traveling towards the 
direction denoted by the three-vector $\vec{v}=(v_x, v_y, v_z)$, we define,
\begin{equation}
    t_x \equiv \begin{cases}
    {\displaystyle\frac{(L-x)}{v_x}}, & \text{if } v_x > 0\,,\\[3.5mm]
    {\displaystyle -\frac{x}{v_x}}, & \text{if } v_x < 0\,,\\
          \end{cases}
\end{equation}
and $t_x > t_y, t_z$ if $v_x=0$, and analogous expressions for $t_y$ and $t_z$, depending on
$W,\,H$, respectively. $L_i$ is then given by,
\begin{equation}
L_i = |\vec{v}| \times \text{min}(t_x, t_y, t_z)\,.
\end{equation}

\paragraph{\texttt{JUNO}:} \texttt{JUNO} has a spherical geometry with a
radius of $R_{\text{max}} = \SI{17.7}{\m}$. For a neutralino produced at 
a random point $\vec r=(r_i,\theta_i, \varphi_i)$ inside the detectable 
region and with velocity $\vec{v}$, flying in the direction given by the 
angles ($\theta_j, \varphi_j$), one can calculate the angle $\theta$ 
between the two vectors: $\cos\theta=\frac{\vec r\cdot\vec v}{|\vec r||\vec v|}$. 
A coordinate transformation is performed in order to eliminate the 
dependence of $\theta$ on $\varphi_i, \varphi_j$, such that $\theta = 
\theta_i-\theta_j\in [0, 2\pi]$. The final distance is calculated to be:
\begin{equation}
L_i = {-r_i\text{cos}(\theta) + \sqrt{R_{\text{max}}^2 - r_i^2 \text{sin}^2(\theta)}}\,.
\end{equation}

\begin{figure}[!htb]
\centering
\begin{minipage}[b]{0.49\textwidth}

\centering
\includegraphics[width= \textwidth]{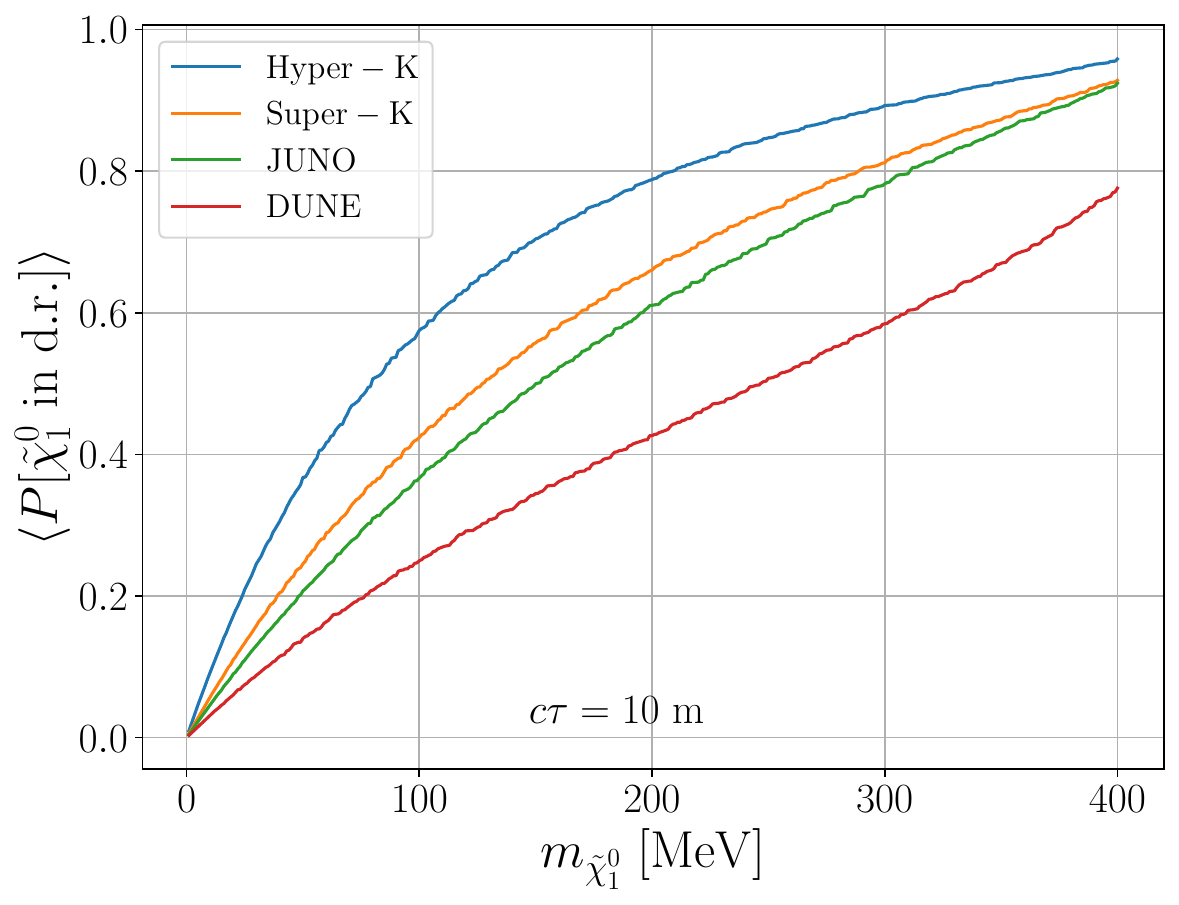}
\end{minipage}
\hfill
\begin{minipage}[b]{0.49\textwidth}
\centering
\includegraphics[width= \textwidth]{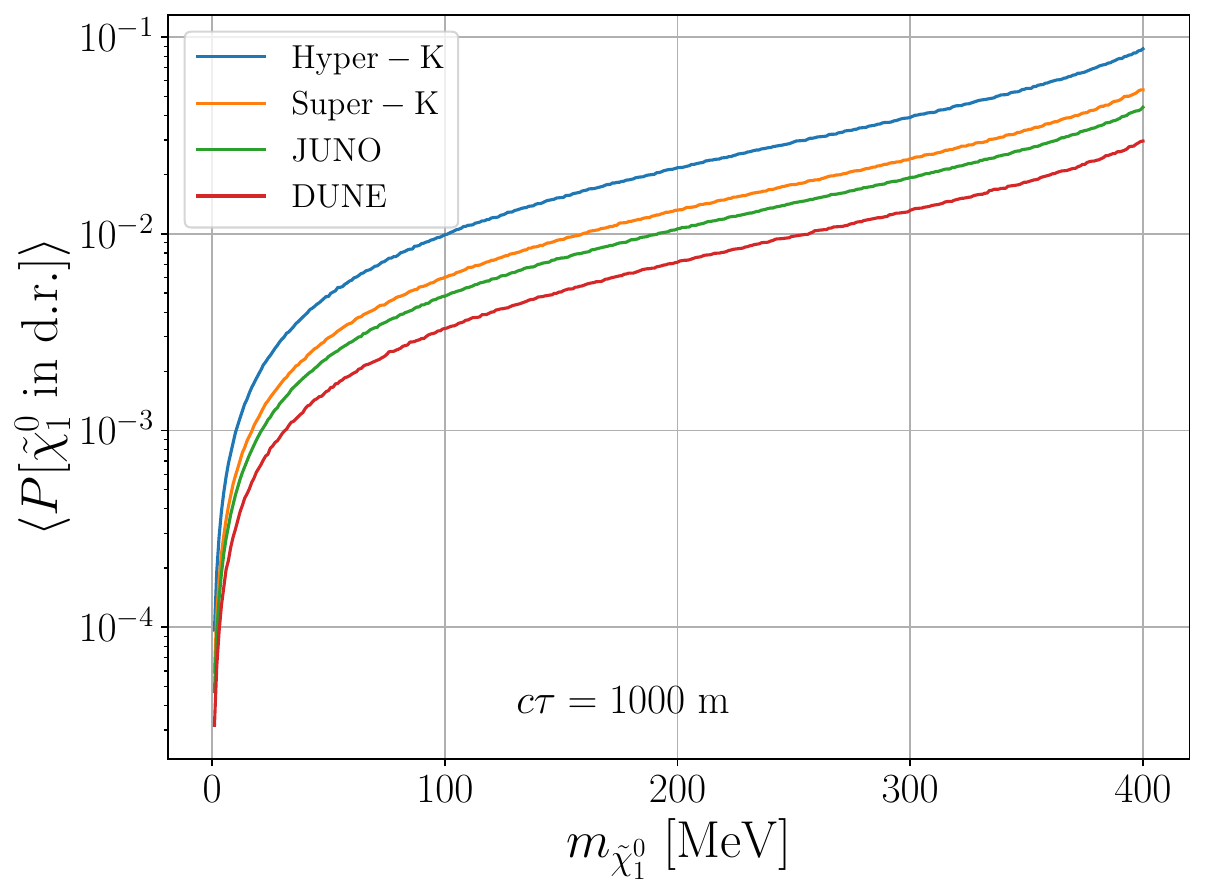}
\end{minipage}

\caption{Average neutralino decay probabilities as a function of
the neutralino mass for fixed neutralino decay length: 
$c\tau= \SI{10}{\m}$ (left) and $c\tau = \SI{1000}{\m}$ (right). 
These plots have been generated with a sample size $N^{\text{MC}}
_{\tilde{\chi}^0_1}= 10,000$.}
\label{fig:avgprob}

\end{figure}

As an illustration, we depict $\langle P[\tilde{\chi}^0_1\ 
\text{in d.r.}]\rangle$ in~\cref{fig:avgprob} as a function of 
the neutralino mass but for fixed decay length, $c\tau$, for the
four detectors. The neutralino momentum $p_{\tilde\chi^0_1}$ at 
production depends on the neutralino mass $m_{\tilde\chi^0_1}$ 
as in Eq.~(\ref{eq:kaonmomentum2}). Thus for increasing $m_{ 
\tilde \chi^0_1}$, $|\vec v_{\tilde\chi^0_1}|$ decreases, and 
the neutralino is more likely to decay in the detector for a fixed lifetime.

%!TEX root= ../RPV_Neutralino_Decay.tex
\section{Numerical Analysis}
\label{sec:Numerical-Analysis}

We now present benchmark scenarios, which, we believe, capture the bulk of 
the phenomenology accessible at \texttt{DUNE}, \texttt{JUNO} and 
\texttt{Hyper-K} for a proton decaying to a lighter neutralino. In all the 
considered cases, the proton disintegration is controlled by the parameter
$\lam''_{121}$, as in~\cref{eq:protonwidth}. Thus, we only consider cases 
for which $m_{\tilde\chi^0_1}\lsim \SI{445}{\mega\electronvolt}$, 
\textit{cf.}~\cref{eq:chimass}. The produced neutralino either 
escapes the detector as missing energy or decays into visible modes via an RPV 
operator $\lam^D_{ijk}$. The benchmarks we study are presented in Section~\ref{subsec:benchmarks} and summarized 
in~\cref{tab:benchmarks}. We present the corresponding numerical studies 
in~\cref{sec:results}. For each scenario, we assume that the listed couplings 
are the only non-negligible RPV couplings; the relevant current bounds are 
also shown in the table.

\subsection{Benchmark Scenarios}\label{sec:scenarios}
\label{subsec:benchmarks}
\begin{table}[ht]
	\begin{adjustbox}{max width=\textwidth}{\begin{tabular}{cccccc}
			\toprule \toprule
			{\bf Scenario} & $\mathbf{m_{\widetilde{\chi}^0_1}}$ & {\bf Proton Decay}
			& $\mathbf{\widetilde{\chi}^0_1}$\;{\bf Decay}$\left(\lambda^D_{ijk}\right)$ & {\bf Product Bound} & {\bf Min.} ${c\tau_{\widetilde{\chi}^0_1}}$\\
                {\bf B1} & $\SIrange{0}{400}{\MeV}$ & $\lambda''_{121}<5\times 10^{-7} \left(\frac{m_{\tilde{q}}}{\tilde{\Lambda}\si{\TeV}}\right)^{5/2}$ & $-$ & $-$ & $\infty$\\
			{\bf B2} & $\SIrange{0}{400}{\MeV}$ & $\lambda''_{121}<5\times 10^{-7} \left(\frac{m_{\tilde{q}}}{\tilde{\Lambda}\si{\TeV}}\right)^{5/2}$ & $\lambda'_{333} <1.04$ & $\lambda'_{333}\lambda''_{121} < 10^{-9}$ & $\sim\SI{1600}{\metre}$\\
			{\bf B3} & $\SIrange{0}{400}{\MeV}$ & $\lambda''_{121}<5\times 10^{-7} \left(\frac{m_{\tilde{q}}}{\tilde{\Lambda}\si{\TeV}}\right)^{5/2}$ & $\lambda_{233}< 0.7\left(\frac{m_{\tilde{\tau}_R}}{\SI{1}{\tera\electronvolt}}\right)$ & $\lambda_{233}\lambda''_{121} < 10^{-21}$  & $\sim\SI{180}{\metre}$\\
			{\bf B4} & $\SIrange{150}{400}{\MeV}$ & $\lambda''_{121}<5\times 10^{-7} \left(\frac{m_{\tilde{q}}}{\tilde{\Lambda}\si{\TeV}}\right)^{5/2}$ & $\lambda'_{211}<0.59\left(\frac{m_{\tilde{d}_R}}{\SI{1}{\tera\electronvolt}}\right)$ & $\lambda'_{211}\lambda''_{121} < 6 \times 10^{-25}$ & $\sim\SI{11}{\metre}$\\
        	{\bf B5} & $\SIrange{150}{400}{\MeV}$ & $\lambda''_{121}<5\times 10^{-7} \left(\frac{m_{\tilde{q}}}{\tilde{\Lambda}\si{\TeV}}\right)^{5/2}$ & $\lambda'_{311}<1.12$ & $\lambda'_{311}\lambda''_{121}<4 \times 10^{-24}$ & $\sim\SI{8}{\metre}$\\
			\bottomrule \bottomrule
	\end{tabular}}
	\end{adjustbox}
	\caption{Details of the benchmark scenarios. The bounds on $\lam^D_{ijk}$ 
are taken from Refs.~\cite{Allanach:1999ic,Dercks:2017lfq} while the one on 
$\lam''_{121}$ is from Ref.~\cite{Super-Kamiokande:2014hie}. Product bounds 
are obtained from Ref.~\cite{Barbier:2004ez} for SUSY masses of $\SI{1}
{\tera\electronvolt}$, except in the case of $\lambda'_{311}\lam''_{121}$, 
where it is the bound on $p \rightarrow K^++\bar{\nu}$ 
from~\cite{Super-Kamiokande:2014otb} reinterpreted for \textbf{B5}. The mass
ranges conform (up to some rounding) to the discussion in the text. 
An additional limit originates from Super-K: $|\lam''_{121}| < 3.9\times 10^{-31} \left(\frac{m_{\tilde{q}}}{\SI{}{\giga\electronvolt}}\right)^2 .$ 
``Min. $c\tau_{\tilde\chi^0_1}$'' refers to the minimal decay length, \textit{i.e.} the decay length at the maximal allowed RPV coupling and neutralino mass within the scenario.}
	\label{tab:benchmarks}
\end{table}

In the first benchmark scenario, \textbf{B1}, we assume that the neutralino 
cannot be observed at \texttt{DUNE}, \texttt{JUNO} or \texttt{Hyper-K}, either 
due to a long lifetime or to invisible decay products. (We set $L$-violating 
couplings to $0$ in practice.)

In the second benchmark, \textbf{B2}, neutralino decays are controlled by the 
trilinear coupling $\lam^{\prime}_{333}$. In this case tree-level decay modes 
[$\tilde\chi_1^0\to (\tau^- t \bar b,\; \nu_\tau b \bar b)$] are kinematically 
prohibited, under the assumption of negligible generation mixing \cite{Dreiner:1991pe, 
Agashe:1995qm}. The radiative channel, mediated by (s)bottom loops, then appears as the dominant one. Existing limits on $\lam^{\prime}_{333}$ imply an already 
sizable decay length (see the last column of~\cref{tab:benchmarks}), as compared to 
the dimensions of the experiments (listed in~\cref{tab:detectorinfo}). We 
briefly comment on alternative choices of dominant $\lam^{\prime}_{ijj}$ 
couplings controlling the radiative decay mode of the neutralino. Due to the 
scaling with the fermion mass squared [see~\cref{eq:neutralinoloop2}], we can 
dismiss the cases $j=1$, $2$ as resulting in very long lifetimes. The choice 
of a dominant $\lam'_{133}$ offers little competition as well, due to severe 
experimental bounds~\cite{Dercks:2017lfq,Allanach:1999ic}: $|\lam'_{133}| < 
1.4\times 10^{-3} \sqrt{\left(\frac{m_{\tilde{q}}}{100\SI{}{\giga\electronvolt}}
\right)}$. $\lam'_{233}$ should perform comparably to $\lambda'_{333}$ 
retained here.

Benchmark \textbf{B3} also involves radiative neutralino decays, now 
controlled by the trilinear coupling $\lam_{233}$ leading to a tau/stau loop.
In this case, significantly shorter decay lengths are accessible 
(see~\cref{tab:benchmarks}), as compared to \textbf{B2}. Other choices of 
$\lam_{ijj}$ would yet lead to very long decay lengths again, if $j=1$, $2$, while 
$\lam_{133}$ is severely constrained experimentally.

The two last scenarios involve tree-level decay modes of the neutralino. Here, 
we dismiss leptonic decays, since the widths associated 
with~\cref{eq:neutralino-decay-LLE} lead to very long-lived neutralinos, 
escaping the detectors in the considered mass regime. Similarly, a very strict 
experimental limit on $\lam'_{111}$~\cite{Dercks:2017lfq} leaves only $\lam' 
_{211}$ and $\lam'_{311}$ in a position to generate semi-leptonic 
disintegrations of the neutralino that are observable at \texttt{DUNE}, 
\texttt{JUNO}, or \texttt{Hyper-K}. For \textbf{B4}, a non-vanishing $\lam'_ 
{211}$ potentially opens the semi-leptonic modes: $\tilde{\chi}^0_1\to \pi^ 
{\pm}+\mu^{\mp}$, $\tilde{\chi}^0_1\rightarrow\pi^{0}+\nu_{\mu},\,\pi^0+\bar
\nu_\mu$, with kinematical limits on the neutralino mass reading $m_{\pi^\pm} 
+ m_{\mu^\mp}\leq m_{\tilde{\chi}^0_1} < m_p-m_{K^+}$ and $m_{\pi^0}\leq m_ 
{\tilde{\chi}^0_1} < m_p-m_{K^+}$, respectively. In benchmark \textbf{B5}, 
$\lam'_{311}\neq0$ triggers the decay channel $\tilde{\chi}^0_1 \rightarrow 
\pi^{0}+\nu_{\tau}$ provided $m_{\pi^0} \leq m_{\tilde{\chi}^0_1}<m_p-m_{K^+}$. 
The SU(2)$_L$-related charged channel is kinematically closed. In both cases, 
the radiative decay mode mediated by a down/sdown loop, although open, results 
in negligible widths due to the Yukawa suppression. 

%!TEX root= ../RPV_Neutralino_Decay.tex
\subsection{Results}\label{sec:results}
We may now discuss the experimental prospects for each of the proposed 
benchmark scenarios. For commodity, we dismiss the comparatively broad 
theoretical uncertainties examplified by~\cref{fig:proton_decay_width_change}, 
\textit{i.e.} work with an absolute prediction of the decay rates, as obtained with the 
hadronic input of Ref.~\cite{Aoki:2017puj}, and strictly focus on a comparative 
analysis of the various experiments in view of identifying a signal of the 
studied kind. This means that the (prospective) bounds on the parameter space of 
the RPV-MSSM shown below should not be understood as absolute, but still 
require a proper account of the theoretical uncertainties before any attempt at 
correlating them with input from other observables.

\begin{figure}
    \centering
    \includegraphics[width=0.7\textwidth]{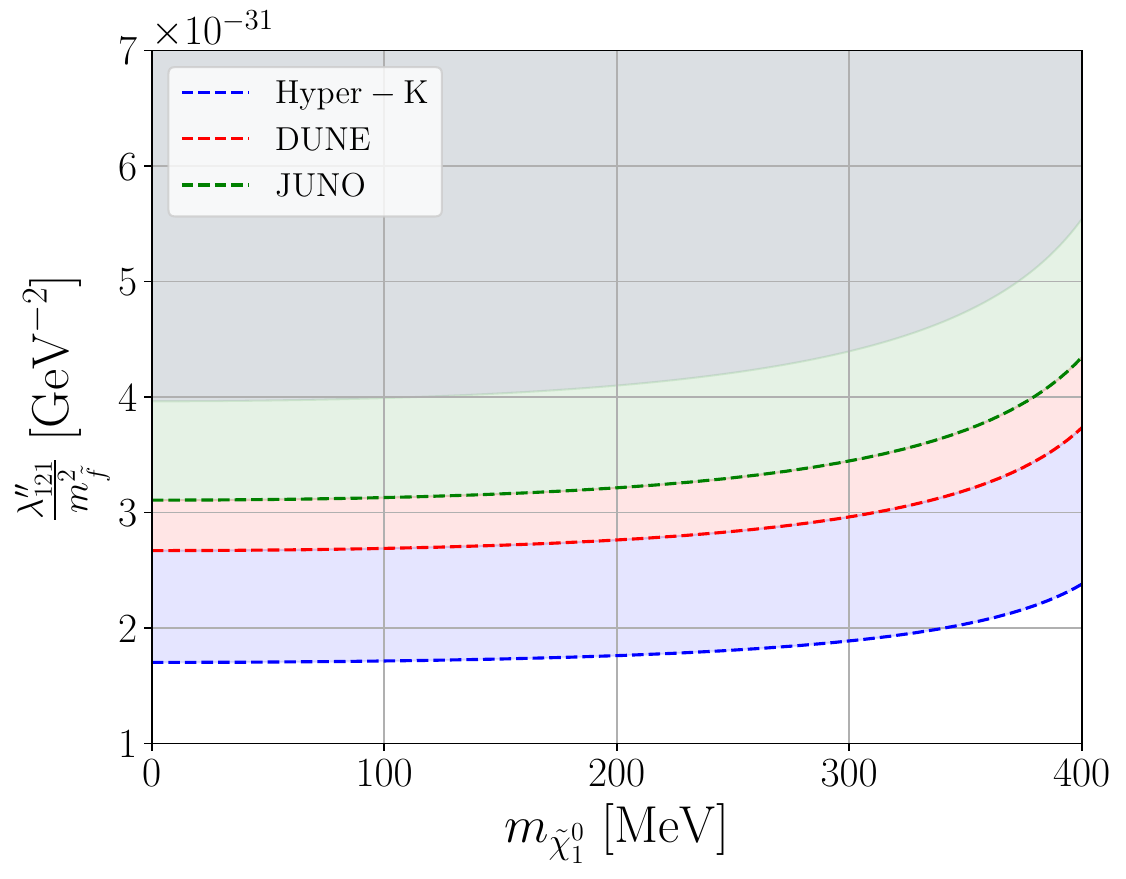}
    \caption{Sensitivity reach for the single coupling scenario of benchmark 
    \textbf{B1}. The reinterpreted bound from \texttt{Super-K} is shown in 
    gray. The bound from~\cref{tab:benchmarks} lies above the scale of the plot. The results for \texttt{Hyper-K}, \texttt{DUNE}, and 
    \texttt{JUNO} are for a run-time of 10 years. }
    \label{fig:B1}
\end{figure}

In the case of \textbf{B1}, with a stable, \textit{i.e.} invisible, neutralino at the 
detector scale, we present in~\cref{fig:B1} the projected parameter space 
coverage achieved by \texttt{DUNE}~\cite{DUNE:2020ypp}, 
\texttt{JUNO}~\cite{JUNO:2022qgr} and 
\texttt{Hyper-K}~\cite{Hyper-Kamiokande:2018ofw} at 90\% confidence level 
sensitivity, after a ten-year run time. Here, we recast the projections quoted 
for the $p\to K^++\bar{\nu}$ mode in the technical reports, under the 
assumption that these remain valid with the modified kaon kinematics. For 
\texttt{Hyper-K}, where the kaon momentum is not measurable, the cases of a 
massive or massless neutralino are indeed indistinguishable\footnote{Most kaons 
($\sim$80\%) decay at rest, and in this case the muons carry no information relative to the 
momentum of their parent kaons.} and our working hypothesis amounts to 
modeling the signal efficiencies and experimental backgrounds as roughly 
constant in the considered kinematical window, as discussed
in~\cref{sec:experiment}. The current \texttt{Super-K} 
bound~\cite{Super-Kamiokande:2014otb} can be exploited in the same fashion, and is included in the plot.
For \texttt{DUNE} and \texttt{JUNO}, however, the searches implement specific 
cuts on the kaon momentum in order to reduce the background due to atmospheric 
neutrinos. This implies that the actual bounds in the massive case would be 
weaker than the naive one, due to eroded signal efficiency. Nevertheless, it 
might be possible to optimize the momentum cuts in searches dedicated to the 
massive neutralino case and counteract this diminished sensitivity. A more 
precise detector simulation goes beyond the scope of the current paper.

\cref{fig:B1} shows the parameter space explored by the 
various experiments in 
the plane spanned by the neutralino mass $m_{\tilde\chi^0_1}$
and $\lam^{\prime\prime}_{121}/m_{ 
\tilde{f}}^2$. The proton decay rate is kinematically suppressed as the
neutralino mass approaches the threshold, $m_{\tilde\chi^0_1}<445\,$MeV; this weakens the reach in $\lam^{\prime\prime}_{121}/
m_{\tilde{f}}^2$ as compared to the massless case. \texttt{JUNO}, \texttt{DUNE} 
and \texttt{Hyper-K} are all expected to improve on the coverage achieved at 
\texttt{Super-K}, probing values of $\lambda^{\prime\prime}_{121}/m_{\tilde{f}}^2$ as much as $1.3$, $1.5$ and $2.5$ smaller, respectively.
Thus, despite \texttt{Hyper-K}'s advantage in terms of fiducial mass, \texttt{JUNO} and \texttt{DUNE} remain competitive 
thanks to their superior efficiencies and high background-rejection rates. In addition, we stress that momentum measurements 
of the produced kaon, needed to correlate the rate with the neutralino mass, will only be possible at these two experiments. While the current level of information presented in technical reports and the difficulty 
of modeling the macroscopic effects leading to momentum smearing make it impossible for us to realistically simulate this search, one should keep in mind that this observable would be simultaneously available 
with the kaon detection search, and might allow \texttt{DUNE} or \texttt{JUNO} to distinguish the scenario with a neutralino from its 
counterpart with a neutrino, even if both these final states result in 
missing energy at colliders.

We now turn to the scenarios with a visible neutralino decay. In this case, the 
relevant parameter space is (at least) three-dimensional, consisting of $m_{ 
\tilde{\chi}_1^0}$, $\lam^{\prime\prime}_{121}/m^2_{\tilde{f}}$ (controlling 
the proton decay, and thus the production rate of the neutralinos),
and 
$\lam^{(\prime)}_{ijk}/m^2_{\tilde{f}}$ (controlling the 
neutralino decay together with the neutralino mass).
For commodity, we focus on two projections in this 
parameter space:
\begin{itemize}
\item First, we examine the plane $\lam^{\prime\prime}_{121}/m^2_{\tilde{f}}$ 
vs.~$\lam^{(\prime)}_{ijk}/m^2_{\tilde{f}}$ at $m_{\tilde{\chi}_1^0}\stackrel{!}
{=}\SI{400}{\MeV}$. 
\item Then, we consider the plane $\lam_{ijk}^{D}/m^2_{\tilde{f}}$ vs.~$m_ 
{\tilde{\chi}^0_1}$, with $\lambda^{\prime\prime}_{121}/m^2_{\tilde{f}}\,$ set to
its value saturating the reinterpreted bound from \texttt{Super-K}.
The corresponding values of $\lambda_{121}^{\prime\prime}/m_{\tilde{f}}^2$ as a function of the neutralino mass would correspond to the edge of the gray area in~\cref{fig:B1} in the case of a purely invisible decay of the neutralino, but are actually shifted when a shorter lifetime makes the invisible search less relevant. 
\end{itemize}
In both cases, we also indicate the neutralino decay lengths, 
as this observable is directly correlated with the considered neutralino decay coupling.
Again, we assume an accumulated ten years of observed data at \texttt{DUNE}, 
\texttt{JUNO} and \texttt{Hyper-K}. As we neglect any sort of background, the 
reach corresponding to 95\% confidence level exclusion is determined by the 
3-event isocurves, which we display in our plots.\footnote{Where relevant, we 
also depict $30$- and $90$-event isocurves.} Existing constraints 
of~\cref{tab:benchmarks} are also depicted in gray (single-bounds) or 
cyan (product-bounds), where relevant.

\begin{figure}[t]
\centering
\begin{minipage}[b]{0.49\textwidth}
\centering
\includegraphics[width= \linewidth]{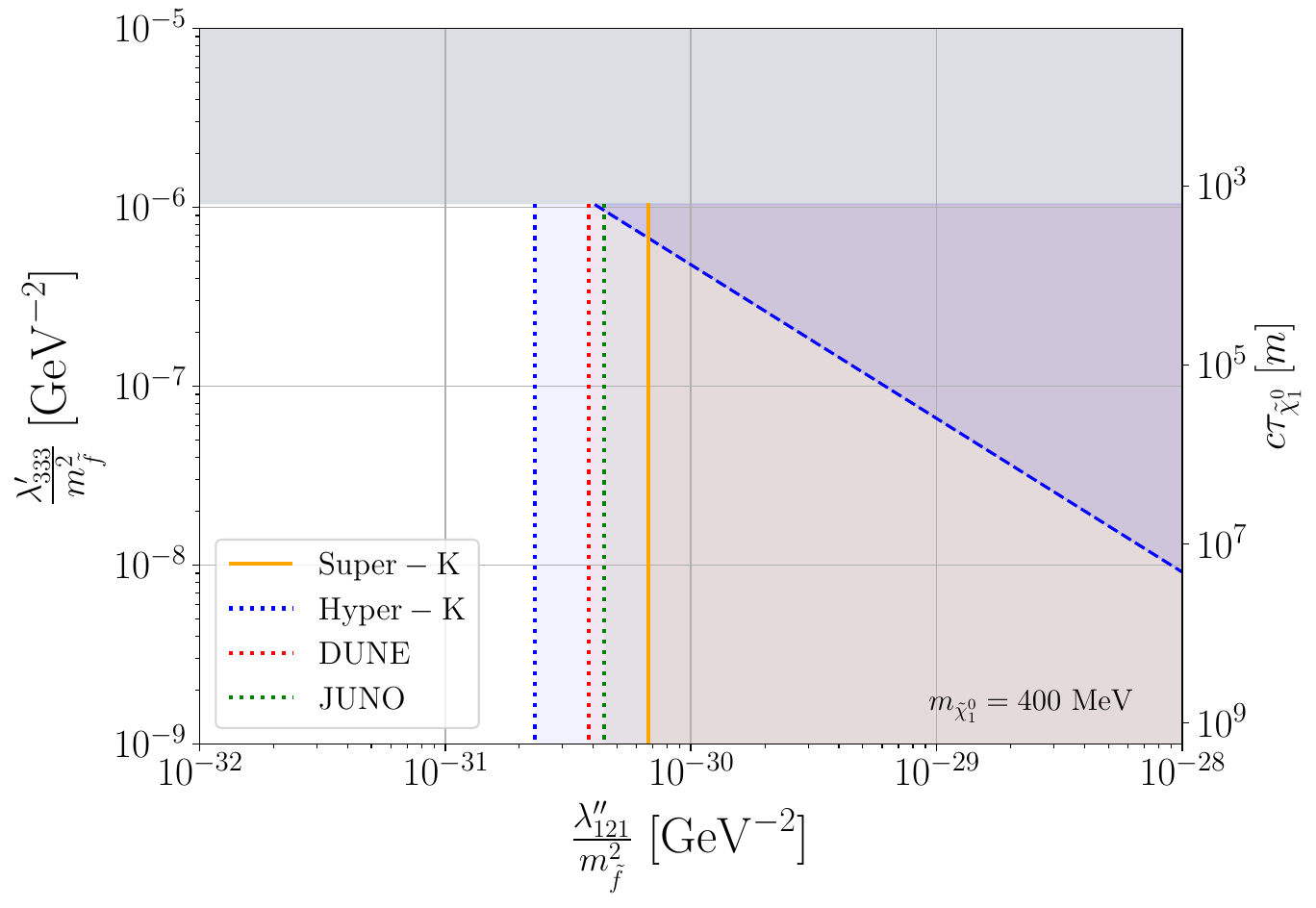}
\end{minipage}
\hfill
\begin{minipage}[b]{0.49\textwidth}
\centering
\includegraphics[width= \linewidth]{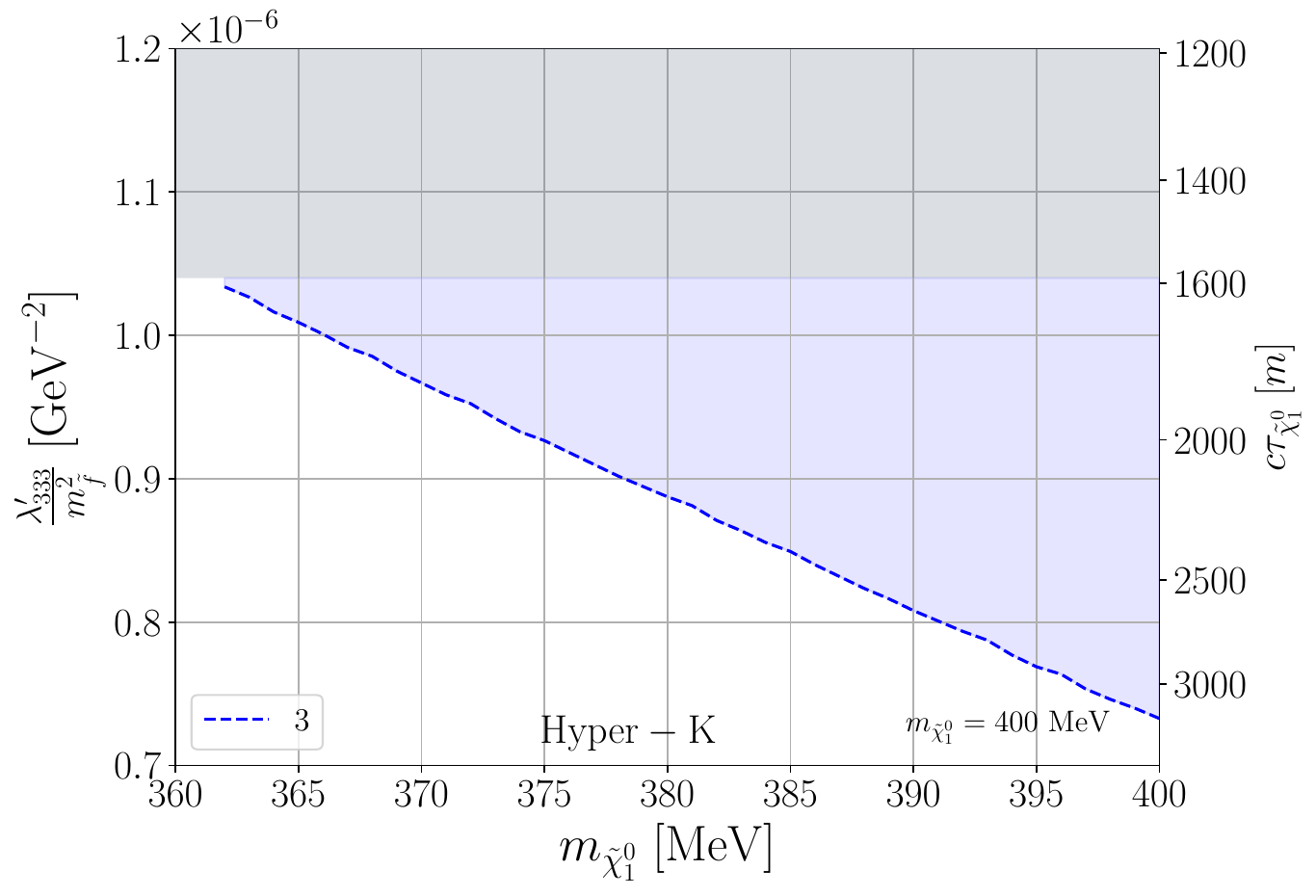}
\end{minipage}
\caption{Sensitivity reach/\texttt{Super-K} limit for benchmark \textbf{B2}. 
The existing single-bound on $\lam'_{333}$ from~\cref{tab:benchmarks} is shown 
in gray (with $m_{\tilde{f}}=\SI{1}{\TeV}$), while the product-bound lies 
outside the scale of the plot. \textit{Left:} The limits in the 
coupling-vs.-coupling plane with $m_{\tilde{\chi}^0_1} = \SI{400}{\MeV}$. The
contours correspond to the visible mode (blue dashed, downward sloping line) 
and invisible mode (vertical lines); see discussion in the text. 
\textit{Right:} The limits in the coupling-vs.-mass plane for the visible mode 
with $\lam_{121}''/m^2_{\tilde{f}}$ fixed at the threshold of the 
\texttt{Super-K} bound of~\cref{fig:B1}.
}

\label{fig:B2}
\end{figure}
We then consider the benchmark~\textbf{B2} with a still long-lived neutralino 
decaying radiatively via bottom\,/\,sbottom loops. The corresponding results are 
shown in~\cref{fig:B2}. Experiments are sensitive to this scenario from 
two directions. First, most of the neutralinos decay outside the detectors, due 
to their long lifetime, hence appear as missing energy, similarly to the scenario of~\cref{fig:B1}. In fact, consulting the right plot
of~\cref{fig:avgprob}, we observe that this happens for over 90\% of the 
neutralinos, on average, even for the shortest available lifetimes and 
a mass as large as $m_{\tilde{\chi}^0_1} = 
\SI{400}{\MeV}$. The corresponding bound hardly depends on $\lam^{\prime}_{333}
/m^2_{\tilde{f}}$ (as long as one is deep in this long-lived regime) and 
produces vertical boundary lines in the plane spanned by $\lam^{\prime\prime}_ 
{121}/m^2_{\tilde{f}}$ and $\lam^{\prime}_{333}/m^2_{\tilde{f}}$, as displayed 
on the left-hand side of~\cref{fig:B2}. Larger values of $\lam^{\prime\prime}_ 
{121}/m^2_{\tilde{f}}\,$ can be 
probed (or, in case of \texttt{Super-K}, are already excluded) by the invisible neutralino search. A 
subsidiary region is probed by the visible search, where only the 3-event line 
associated with \texttt{Hyper-K} appears on the plot (the corresponding 
boundaries for \texttt{DUNE} and \texttt{JUNO} would be far to the right). 
We should stress, here, that our estimates for the efficiencies achieved in the visible search are very optimistic and that the reach of this observable may be significantly more reduced than what is presented in~\cref{fig:B1}. The 
number of visible decays scales like $|\lam'_{333}\lam_{121}''/m^4_{\tilde{f}}|^2$
in the limit of long neutralino lifetimes, resulting in a sloping coverage 
limit. The large volume of \texttt{Hyper-K} is a clear advantage for detection sensitivity.

\begin{figure}[t]
\centering
\begin{minipage}[b]{0.49\textwidth}
\centering
\includegraphics[width= \textwidth]{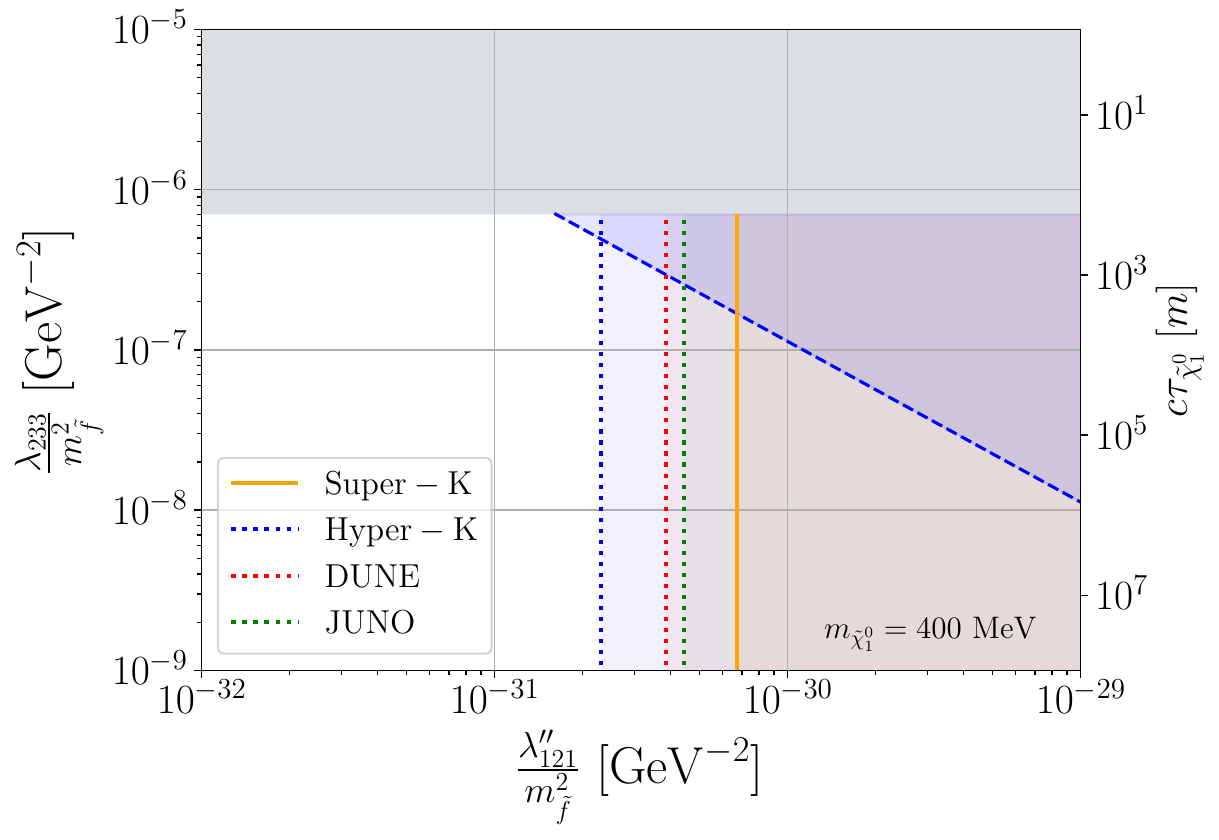}
\end{minipage}
\hfill
% \begin{minipage}[b]{0.49\textwidth}
% \centering
% \includegraphics[width= \textwidth]{figs/HKcase2B2.pdf}
% \end{minipage}
%\hfill
\begin{minipage}[b]{0.49\textwidth}
\centering
\includegraphics[width= \textwidth]{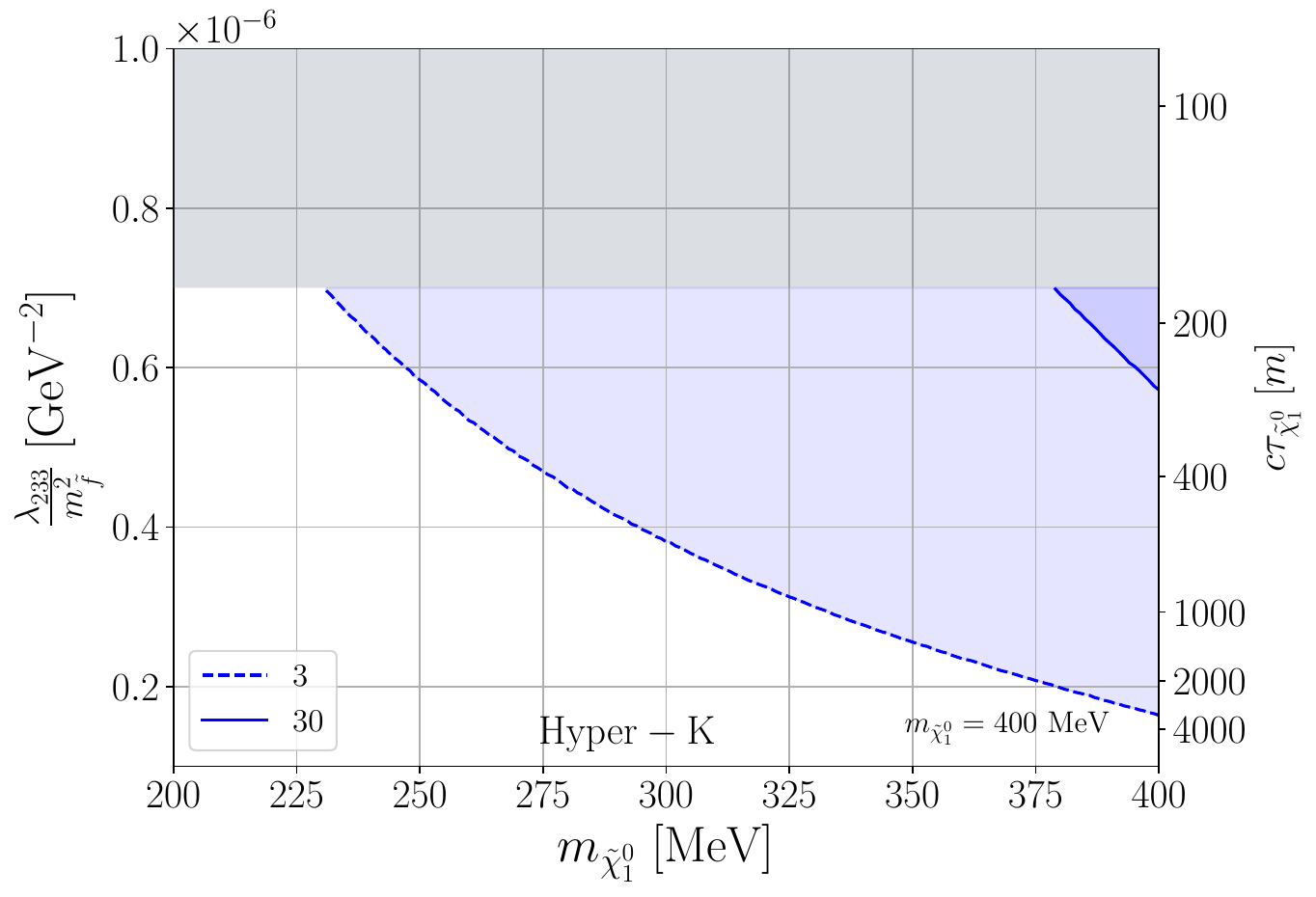}
\end{minipage}
\caption{Sensitivity reach/\texttt{Super-K} limit for benchmark \textbf{B3}. The existing single-bounds from~\cref{tab:benchmarks} are shown in gray while the product-bound is shown in blue (all with $m_{\tilde{f}}=\SI{1}{\TeV}$). \textit{Top Left:} As in left plot of~\cref{fig:B2} but for benchmark \textbf{B3}. \textit{Top Right:} Zoomed-out version of the top-left plot. \textit{Bottom:} As in right plot of~\cref{fig:B1} but for benchmark \textbf{B3}. The dashed and solid lines correspond to $3$- and $30$-event isocurves, respectively.
}
\label{fig:B3}
\end{figure}

In the right plot of the figure, we show the sensitivity reach of the visible 
mode in the $\lam'_{333}/m^2_{\tilde{f}}$ vs. $m_{\tilde{\chi}^0_1}$ plane, 
where $\lam_{121}^{\prime\prime}/m_{\tilde{f}}^2$ is always set at the exclusion 
limit of \texttt{Super-K}.\footnote{All the corresponding points are thus 
simultaneously, and independently, probed by the invisible search at 
\texttt{Hyper-K}, \texttt{DUNE}, and \texttt{JUNO}.} The 
sensitivity improves as the neutralino mass increases due to (a) the lifetime 
becoming shorter, \textit{cf.}~\cref{eq:neutralinoloop2}, and (b) the 
neutralinos having lower momentum: the decay length of the 
neutralino $(\beta\gamma c\tau)_{\tilde\chi^0_1}$ is indeed shorter, resulting 
in an increased average decay sensitivity $\langle P[\tilde{\chi}^0_1\ 
\text{in d.r.}]\rangle$, as shown in~\cref{fig:avgprob}.

The physics situation of benchmark~\textbf{B3} is largely comparable to 
that of the former benchmark, with a radiatively decaying neutralino. 
Consequently, the parameter space coverage in proton decay experiments 
follows the same trends, as can be observed in~\cref{fig:B3}. Shorter 
lifetimes are accessible nonetheless, resulting in an increased relevance 
of the visible search channel, although only \texttt{Hyper-K} has viable 
detection prospects in this mode. The left-hand plot of~\cref{fig:B3} 
distinctly exhibits a region in the higher range of $\lambda^{\prime}_{233}
/m^2_{\tilde{f}}$ where both types of detection are competitive, as well 
as a tiny region only accessible to the visible search. Once again, we 
point at the generous efficiencies assumed for the visible search here: 
this region accessible only to the visible search would likely shrink with 
more realistic estimates.

% The results for benchmark \textbf{B3}, also corresponding to a radiative decay, are 
% very similar to benchmark \textbf{B2}; the coupling-vs.-coupling (top) and 
% coupling-vs.-mass (bottom) plots are shown in~\cref{fig:B3}. 
% %
% %
% Once again, \texttt{Hyper-K} is the only relevant experiment for detecting the 
% visible mode. This time, the mode has a higher sensitivity owing to the shorter 
% lifetime of the neutralino here. In the top-left
% plot we see, that again, there is a significant parameter range, where both modes 
% are detectable. This could lead to a disentangling of the underlying mechanism
% for proton decay. Furthermore, there is a small region of parameter space, where
% the visible mode is \textit{the only} detectable signature. In the top-right plot, 
% we show a zoomed-out version of the same plot for context. Interestingly, vast 
% regions of parameter space beyond the bounds of~\cref{tab:benchmarks} are already 
% ruled out by the reinterpreted limit from \texttt{Super-K}, even in a scenario 
% where the neutralino is unstable. The experiments at \texttt{Hyper-K}, \texttt{DUNE},
% and \texttt{JUNO} are sensitive slightly beyond the \texttt{Super-K} bound, similar to~\cref{fig:B1}.
% %
% The coupling-vs.-mass plot (bottom) shares similar features to before. This 
% time, we also depict the $30$-event isocurve.

\begin{figure}[!htb]
\centering
\begin{minipage}[b]{0.49\textwidth}
\centering
\includegraphics[width= \textwidth]{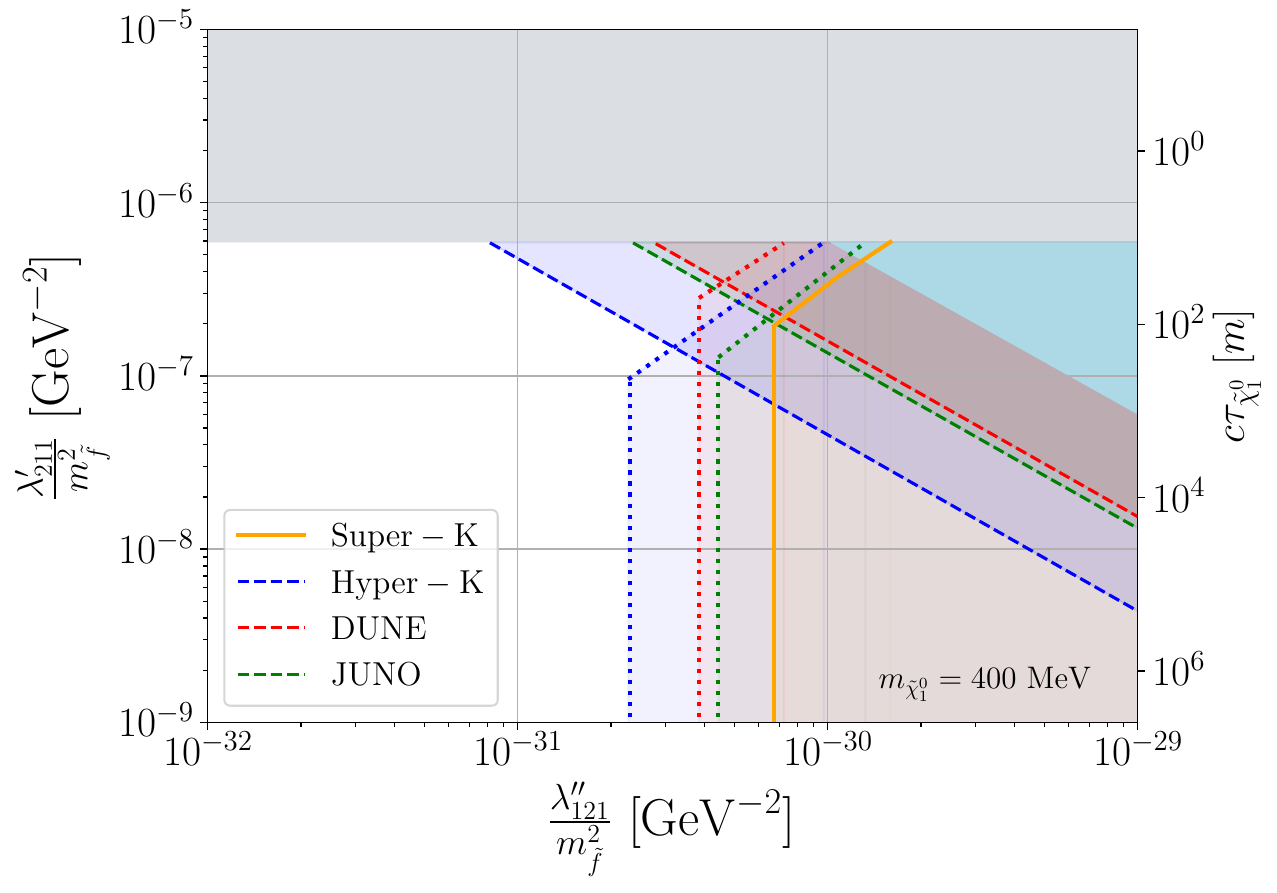}
\end{minipage}
\hfill
\begin{minipage}[b]{0.49\textwidth}
\centering
\includegraphics[width= \textwidth]{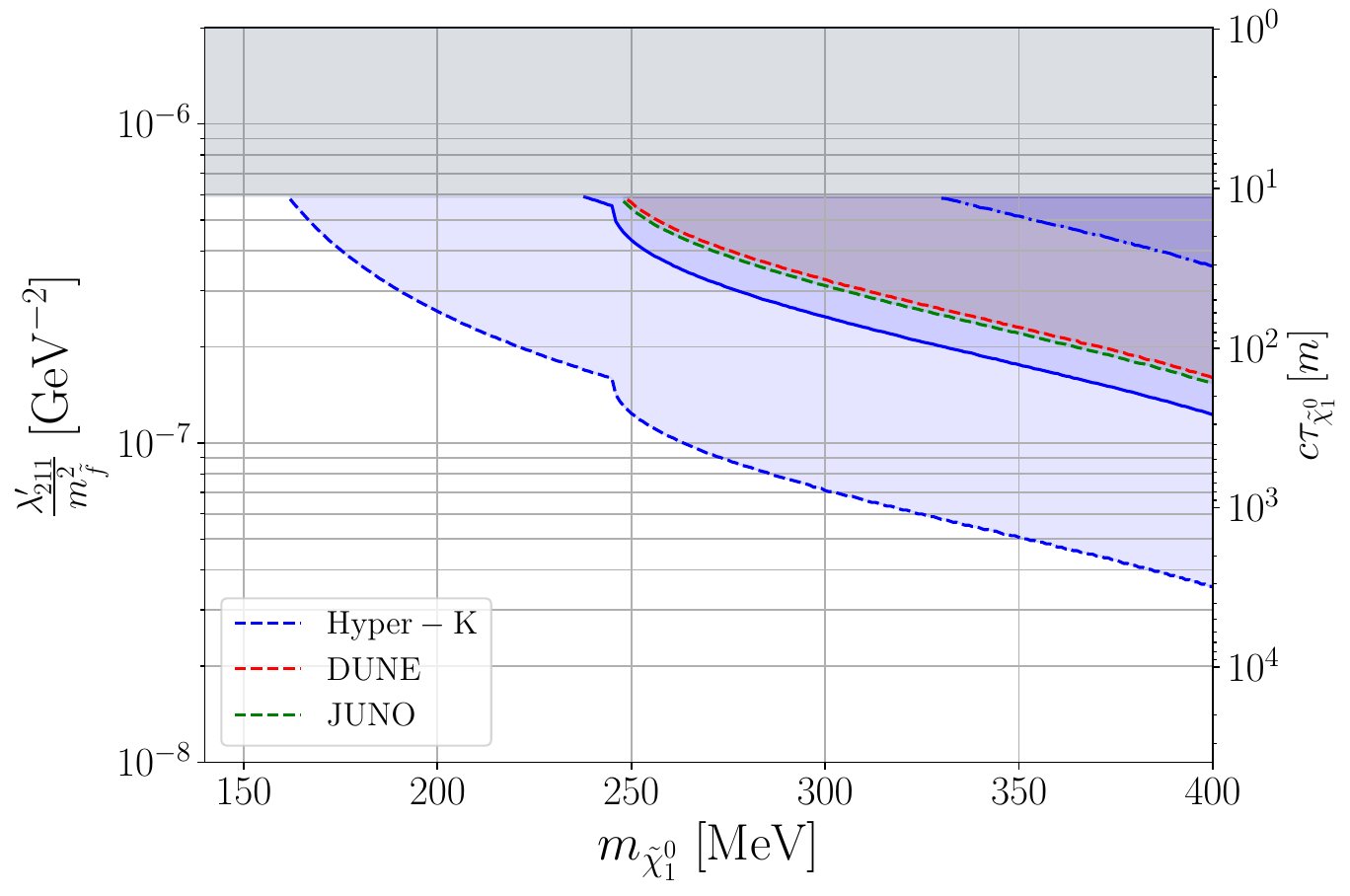}
\end{minipage}
\caption{Sensitivity reach/\texttt{Super-K} limit for benchmark \textbf{B4}. The existing single-bound on $\lambda'_{211}$ from~\cref{tab:benchmarks} is shown in gray, the product-bound is in light blue (both with $m_{\tilde{f}}=\SI{1}{\TeV}$), while the bound on $\lambda''_{121}$ lies outside the scale of the plot. \textit{Left:} As in left plot of~\cref{fig:B2} but for benchmark \textbf{B4}. \textit{Right:} As in right plot of~\cref{fig:B1} but for benchmark \textbf{B4}. The dashed, solid, and dot-dashed lines correspond to $3$-, $30$- and $90$-event isocurves, respectively. An interesting thing to note is the kink in sensitivity in the right figure around $m_
{\tilde{\chi}^0_1} \sim \SI{240}{ \MeV}$, which is due to the modes $\tilde{\chi}^0_1 
\to \pi^{\pm}+\mu^{\mp}$ being kinematically allowed, thus increasing the total decay width.}
\label{fig:B4}
\end{figure}

\begin{figure}[!htb]
\centering
\hfill
\begin{minipage}[b]{0.49\textwidth}
\centering
\includegraphics[width= \textwidth]{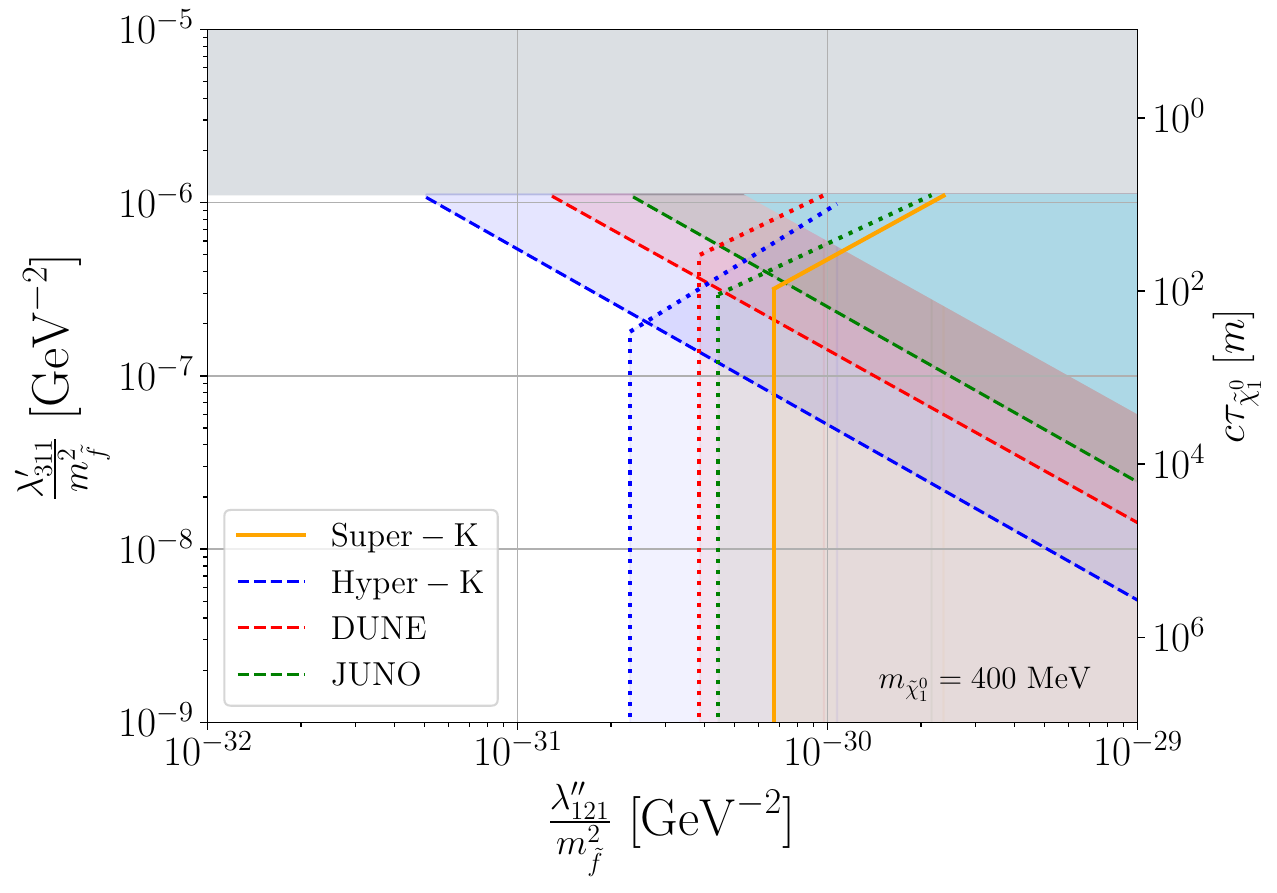}
\end{minipage}
\hfill
\begin{minipage}[b]{0.49\textwidth}
\centering
\includegraphics[width= \textwidth]{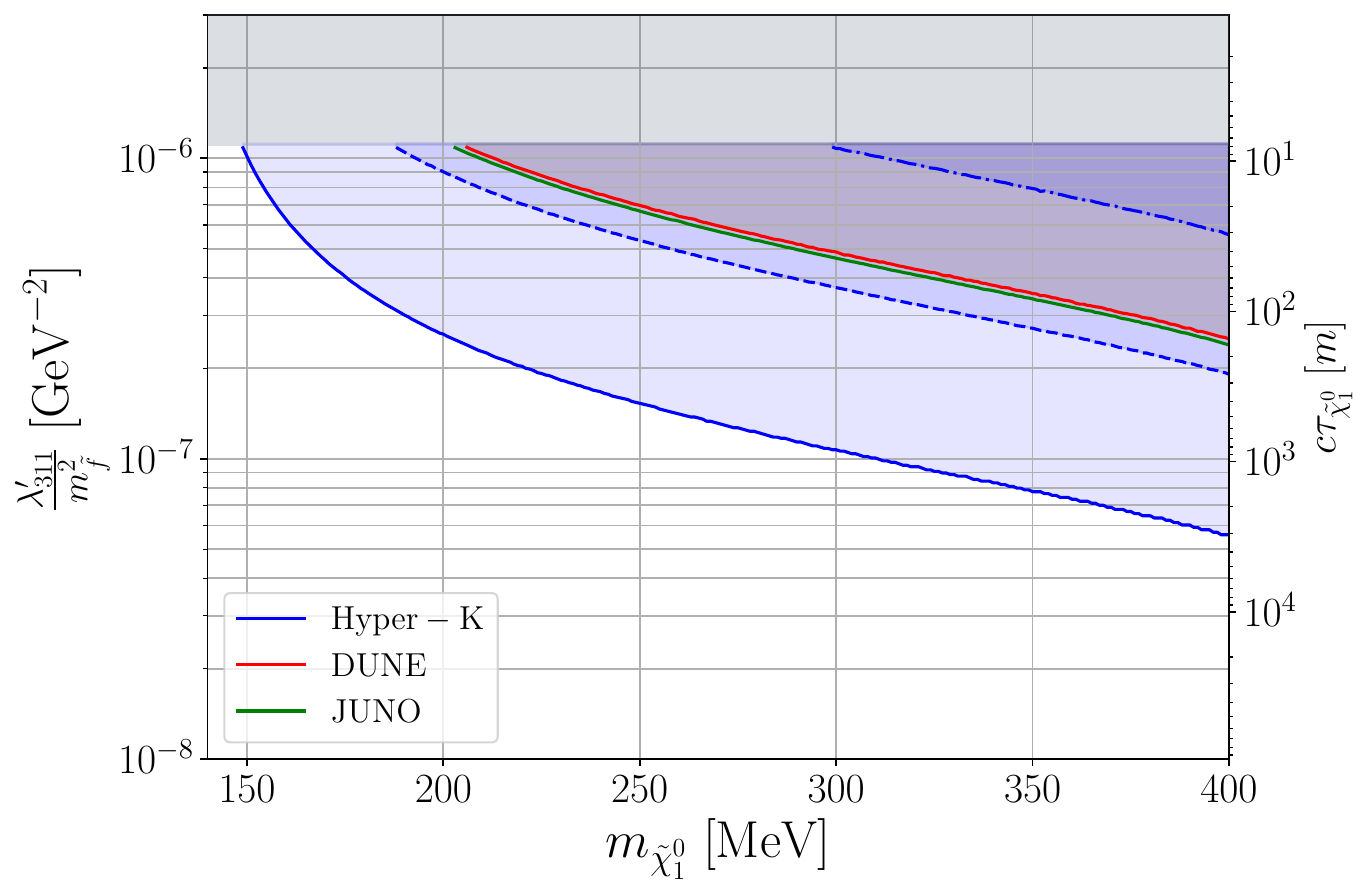}
\end{minipage}
\caption{Sensitivity reach/\texttt{Super-K} limit for benchmark \textbf{B5}. The existing single-bound on $\lambda'_{311}$ from~\cref{tab:benchmarks} is shown in gray, the product-bound is in light blue (both with $m_{\tilde{f}}=\SI{1}{\TeV}$), while the bound on $\lambda''_{121}$ lies outside the scale of the plot. \textit{Left:} As in left plot of~\cref{fig:B2} but for benchmark \textbf{B5}. \textit{Right:} As in right plot of~\cref{fig:B1} but for benchmark \textbf{B5}. The dashed, solid, and dot-dashed lines correspond to $3$-, $30$- and $90$-event isocurves, respectively.}
\label{fig:B5}
\end{figure}

In the case of benchmarks \textbf{B4} and \textbf{B5}, the tree-level decay modes of the neutralino may result in 
comparatively short lifetimes when the RPV couplings are set to their maximal values compatible with existing 
bounds. As a consequence, the visible search strategy is expected to perform more competitively than for the previous
scenarios. Our results are shown in~\cref{fig:B4} and~\cref{fig:B5}. There, a portion of the parameter space left 
open by \texttt{Super-K} is accessible to visible searches at \texttt{DUNE} or \texttt{JUNO}, although 
\texttt{Hyper-K} remains the experiment most sensitive to this mode (as well as to the invisible search).
Conversely, the invisible search channel becomes less competitive in the higher range of $\lam'_{211}/m^2_{\tilde f}$
due to fewer neutralinos escaping the detectors: this effect is observable as a kink in the boundaries, which transit 
from the $\lam'_{i11}/m^2_{\tilde f}$-blind regime at $\lam'_{211}/m^2_{\tilde f}\lsim 10^{-7}$ to a power-law 
regime (appearing as a line in logarithmic scales). Interestingly, \texttt{DUNE} proves slightly more performant than 
\texttt{Hyper-K} in the upper range of $\lam'_{i11}/m^2_{\tilde f}$ for the invisible search.

% For benchmarks \textbf{B4} and \textbf{B5}, the neutralino has tree-level decays leading 
% for maximal values of the couplings to lifetimes significantly shorter than the other 
% benchmarks. The results are shown in~\cref{fig:B4} and~\cref{fig:B5}, respectively. This 
% time, \texttt{DUNE} and \texttt{JUNO} also show sensitivity to the visible mode. Further, 
% here we have a significant parameter range, with sensitivity well below the 
% \texttt{Super-K} bound on $\lam''_{121}$. The other features are as before. The vertical lines 
% that \sout{start sloping upwards} \HKD{(vertical lines can not slope upwards, they are already vertical)
% have a kink at $\lam'_{211}/m^2_{\tilde f}\gsim 10^{-7}$ and then rise linearly in the logarithmic plot
% as a function of $\lam'_{i21}/m^2_{\tilde f}$} 
% are the contours from the missing-energy search. The shape can be understood in the following way. As $\lam'_{i11}/m^2_{\tilde{f}}$ 
% increases, more neutralinos decay inside the detector, reducing the 
% sensitivity of the invisible channel. For $\lam_{121}''/m^2_{\tilde{f}}$ values
% lying below the thresholds
% of \cref{fig:B1}, the number of produced neutralinos is too low, and there is no sensitivity.
% One interesting thing to note is the kink in sensitivity in the right figure around $m_
% {\tilde{\chi}^0_1} \sim \SI{240}{ \MeV}$, which is due to the modes $\tilde{\chi}^0_1 
% \to \pi^{\pm}+\mu^{\mp}$ being kinematically allowed, thus increasing the total decay width.

To summarize this discussion, we observed that the invisible search strategy remains highly relevant for long-lived 
neutralinos produced in proton decays, with an evident advantage in regions of the parameter space where neutralino
decays are suppressed, \textit{i.e.}~in the lower range of the L-violating trilinear couplings or in the regime dominated by 
radiative decays. There, \texttt{DUNE}, \texttt{JUNO} and \texttt{Hyper-K} would typically improve on \texttt{Super-K} by 
a factor 2. We nonetheless stress that this performance depends on the validity of the reinterpretation of the 
searched massless neutrino in terms of a massive neutralino. Here, the measurement of the kaon momentum, possible at
\texttt{DUNE} or \texttt{JUNO} in principle, could help discriminate between the two cases and hint at a neutralino-like 
scenario. On the other hand, the visible search channel allows for a complementary and largely independent coverage of 
parameter space when L-violating effects are sufficiently large to enhance the number of neutralinos decaying in the detectors. 
Such a signal would be a clear indication for a proton decay mode beyond the straightforward (and possibly also supersymmetric) $K^++\bar{\nu}$ 
channel. In a narrow region of the RPV-MSSM parameter space, both visible and invisible strategies could be simultaneously 
successful and result in observations at one (not necessarily the same) of the three planned experiments. A statistical 
combination of the two signals would then be possible, leading to an expectedly even higher sensitivity. Finally, we wish 
to stress the impact of one of our working assumptions: with more than one non-zero L-violating couplings, multiple decay 
channels would open for the neutralino, leading to an increased relevance of the visible over the invisible search strategy.

% \begin{figure}
% \centering
% \begin{subfigure}{0.45\textwidth}
% \centering
% \includegraphics[width=\textwidth]{figs/HK_stable.pdf}
% \end{subfigure}\hfill
% \begin{subfigure}{0.45\textwidth}
% \centering
% \includegraphics[width= \textwidth]{figs/DUNE_stable.pdf}
% \end{subfigure}\hfill
% \begin{subfigure}{0.45\textwidth}
% \centering
% \includegraphics[width= \textwidth]{figs/JUNO_stable.pdf}
% \end{subfigure}

% \caption{\textbf{B5}: Sensitivity reach for $\lambda''_{121}/m_{\tilde{f}}^2$ for \texttt{Hyper-K}, \texttt{DUNE} and \texttt{JUNO} as a function of the neutralino mass $m_{\tilde{\chi}^0_1}$. The dashed lines show 3 event isocurves. The bound obtained from \texttt{Super-K} is shown in gray.
% \DK{@Apoorva. Fix label. See above!  Please increase label font!}}
% \label{fig:B5}

% \end{figure}

%\AS{We find that in all five benchmark scenarios, sensitivity beyond the current RPV upper bounds can be achieved. For single bounds, a factor of $\sim 2$ improvement is achieved for the best case scenario, when $m_{\tilde{\chi}^0_1} \sim  \SI{400}{\MeV}$. For products bounds, improvement in order of $\mathcal{O}(10^{-6})$ or higher is achieved}

%!TEX root= ../RPV_Neutralino_Decay.tex
\section{Conclusions}\label{sec:conclusions}
The search for proton decay is tightly related to the fundamental and 
experimentally well-established question of matter-antimatter asymmetry. 
Experimental and phenomenological activity in this respect thus appears 
strongly motivated. In this work, we tried to emphasize a less standard 
but very realistic decay mode of the proton, $p\to K^++\tilde{\chi}_1^0$. 
Here $\tilde{\chi}_1^0$ denotes a light and long-lived exotic neutral 
particle, e.g.~a bino-like neutralino in the RPV-MSSM, which can possibly 
decay within the detector. We demonstrate the strong potential of the 
upcoming \texttt{DUNE}, \texttt{JUNO} and \texttt{Hyper-K} experiments to 
investigate such a scenario. This analysis was performed under the 
assumption that the experimental searches considering a massless neutrino 
can be reinterpreted in terms of a massive exotic particle, or that 
experimental cuts can be adjusted in order to address this situation. 
Several signatures can be looked for, depending on the lifetime of the 
neutralino and its decay channels. Measurement of the kaon momentum (at 
\texttt{DUNE} or \texttt{JUNO}) or observation of a displaced vertex from 
the neutralino decay would provide means to distinguish such a scenario 
from the more traditionally considered $p\to K^++\bar{\nu}$. We have 
illustrated these features with the help of a collection of benchmarks. A 
significant coverage of the parameter space can be achieved by \texttt{
DUNE}, \texttt{JUNO} and, most especially, \texttt{Hyper-K}, clearly 
improving on the current limits obtained at \texttt{Super-K}. More 
detailed collider simulations would be necessary to precisely delimit the 
reach of these experiments. In any case, we stress the necessity to keep 
the scope of the experiments as broad as possible and allow the 
reinterpretation of results in non-standard scenarios.

\appendix

\bigskip
\section*{Acknowledgments}
\bigskip
We thank Artur Ankowski, Marcela Carena, Felix Kling, Ornella Palamara,  Rob Timmermans,
Simon Zeren Wang, 
and Tingjun Yang for useful discussions. HKD thanks the Nikhef theory group for
its hospitality while part of this work was completed. This work was supported by
the Deutsche Forschungsgemeinschaft (DFG, Ger- man Research Foundation) through 
the funds provided to the Sino-German Collaborative Research Center TRR110 
“Symmetries and the Emergence of Structure in QCD” (DFG Project ID 196253076 - 
TRR 110).

\appendix
\section{Heavy Neutral Lepton}
\label{app:HNL}

Throughout our work we have focussed on decays to and of a supersymmetric 
neutralino in a model with broken R-parity. 
As explained in~\cref{sec:model}, the low-energy (proton-scale) EFT's of 
Refs.~\cite{Chamoun:2020aft, Domingo:2022emr} provide a natural framework 
to study such processes. The imprint of the high-energy (in our case SUSY) 
model is then encoded within the Wilson coefficients, determined by the 
matching procedure and the Renormalization Group Equations. In this 
fashion, we were always able to present our results in terms of the 
parameters of the more fundamental model, namely the neutralino mass 
$m_{\tilde\chi^0_1}$, the RPV couplings $\lam  
''_{ijk},\, \lam'_{ijk},\,\lam_{ijk}$, and the universal scalar 
fermion mass $m_{\tilde f}$. In the production of the 
neutralino via proton decay baryon-number is violated. In the 
decay of the neutralino lepton-number is violated. Here, we have 
associated no lepton- or baryon- number with the (Majorana) 
neutralino. 

As outlined in the introduction a right-handed neutrino, $N_R$, potentially 
light, has the same SM gauge quantum numbers as the light (dominantly bino) 
neutralino. Associating the lepton number 1 with such a Majorana particle,
$N_R$,
thus appears as a largely formal distinction in the L-violating context and 
the expected phenomenology in proton disintegrations is formally unchanged. 
In particular, the proton-scale EFT's of Refs.~\cite{Chamoun:2020aft, 
Domingo:2022emr} (with the neutralino replaced by $N_R$) remain fully 
operational intermediaries between high-energy scales and hadronic physics. 
Differences of the UV completion in the neutralino case (RPV-SUSY) and any potential UV completion for the HNL model will emerge at the level of the matching.

The phenomenology of the right-handed neutrino can be studied in the context 
of $\nu$SMEFT~\cite{delAguila:2008ir, Liao:2016qyd, Bischer:2019ttk}, itself 
a low-energy EFT valid at the electroweak scale. Obvious contributions to 
the proton-decay operators are generated by the operators listed in Eq.~(7) 
in Ref.~\cite{delAguila:2008ir}. We insist upon the fact that these 
contributions from $\nu$SMEFT potentially involve low-energy operators beyond 
$\hat{\mathcal{Q}}'_1,\,\hat{\mathcal{Q}}_1,\,\hat{\mathcal{Q}}_2$ of 
Eq.~(\ref{eq:chi_operators2}). Indeed, SUSY (through the holomorphicity 
condition of the superpotential) favors baryon-number violation involving 
right-handed [\textit{i.e.}~SU(2)-singlet] quarks, while such a prejudice need not 
apply in $\nu$SMEFT.\footnote{Even in the RPV-MSSM, 
contributions restrict to the operators $\hat{\mathcal{Q}}'_1,\,\hat{\mathcal{Q}}_1,
\,\hat{\mathcal{Q}}_2$ only under simplifying assumptions regarding 
\textit{e.g.}~the mixing among SUSY particles.}

Similarly, the $\nu$SMEFT operators listed in Eq.~(6) in Ref.~\cite{delAguila:2008ir}
and denoted $\mathcal{O}_{LNLe}$, $\mathcal{O}_{LNQd}$, and 
$\mathcal{O}_{QNLd}$ straightforwardly contribute to the decays of the right-handed neutrino. 
Corresponding effects can be projected onto the low-energy operator list presented in detail 
in Ref.~\cite{Domingo:2022emr}, in the context of a light neutralino.

%%%%%%%%%%%%%%%%%%%%%%%%%%%%%%%%%%%%%%%%%%%%%%%%%%%%%%%%%%%%%%%%%%%%%%
\bibliographystyle{JHEP}

\bibliography{bibliography}
\end{document}